\documentstyle[11pt,aaspp]{article} 
\newlength{\parindnt}
\def\gapprox{\;\rlap{\lower 2.5pt
 \hbox{$\sim$}}\raise 1.5pt\hbox{$>$}\;}
\def\gsim{\;\rlap{\lower 2.5pt
 \hbox{$\sim$}}\raise 1.5pt\hbox{$>$}\;}
\def\lapprox{\;\rlap{\lower 2.5pt
   \hbox{$\sim$}}\raise 1.5pt\hbox{$<$}\;}
\def\lsim{\;\rlap{\lower 2.5pt
   \hbox{$\sim$}}\raise 1.5pt\hbox{$<$}\;}

\newcommand\beq{\begin{equation}}
\newcommand\eeq{\end{equation}}
\newcommand{\Ref}{\hangindent=20pt \hangafter=1 \noindent}
\newcommand{\StartRef}{\hyphenpenalty=10000 \raggedright
\parskip=0pt \parindent=0pt }

\def\lya{Ly$\alpha$}

\begin{document}

\title{Signatures of Stellar Reionization of the Universe}
\author{Zolt\'an Haiman and Abraham Loeb}
\medskip
\affil{Astronomy Department, Harvard University, Cambridge, MA 02138}

\begin{abstract}

The high ionization level and non-zero metallicity ($\sim
1\%~Z_\odot$) of the intergalactic gas at redshifts $z\la5$ implies
that nonlinear structure had started to form in the universe at
earlier times than we currently probe.  In Cold Dark Matter (CDM)
cosmologies, the first generation of baryonic objects emerges at
redshifts $z\sim 10$--$50$.  Here we examine the observable
consequences of the possibility that an early generation of stars
reionized the universe and resulted in the observed metallicity of the
\lya~forest.  Forthcoming microwave anisotropy experiments will be
sensitive to the damping of anisotropies caused by scattering off free
electrons from the reionization epoch.  For a large range of CDM
models with a Scalo stellar mass function, we find that reionization
occurs at a redshift $z\ga 10$ and damps the amplitude of anisotropies
on angular scales $\la 10^\circ$ by a detectable amount, $\sim
10$--$25\%$. However, reionization is substantially delayed if the
initial stellar mass function transformed most of the baryons into low
mass stars. In this case, the mass fraction of pre--galactic stars
could be constrained from the statistics of microlensing events in
galactic halos or along lines of sight to quasars.  Deep infrared
imaging with future space telescopes (such as SIRTF or the Next
Generation Space Telescope) will be able to detect bright star
clusters at $z\ga 5$. The cumulative Bremsstrahlung emission from
these star clusters yields a measurable distortion to the spectrum of
the microwave background.

\noindent
\keywords{cosmology:theory--galaxies:formation--molecular processes--radiative transfer}

\end{abstract}

\section{Introduction}

Currently, a number of competing models for structure formation in the
universe have comparable success in accounting for the observed large scale
structure at $z\lsim 5$ (based on galaxy surveys and quasar absorption line
studies) and at $z\sim10^3$ (based on microwave anisotropy experiments).
Thus, an efficient method for identifying the unique model that describes
structure formation in the real universe is to study the early collapse of
small mass objects at the intermediate redshift interval $5\la z\la10^3$.
Variants on a popular cosmology which make similar predictions about the
formation of structure on large scales, might differ {\it qualitatively} in
their predictions about structure on small scales.  For example, the
standard Cold Dark Matter (CDM) power--spectrum of fluctuations predicts a
bottom--up hierarchy of structure formation which is only mildly
(logarithmically) satisfied at the low mass end of the fluctuation
spectrum. Any power--law tilt of this spectrum that adds more power on
large scales would change the hierarchy of low--mass objects to that of a
top--down scenario, reminiscent of hot dark matter.

In CDM cosmologies, the first baryonic objects with masses $\sim
10^5M_\odot$ form at a redshift as high as $z\sim 50$ (see Haiman,
Thoul \& Loeb 1996, and references therein). It has long been realized
that star formation in these objects could release a sufficient flux
of UV photons to reionize the universe (Couchman \& Rees 1986;
Fukugita \& Kawasaki 1994; Shapiro, Giroux, \& Babul 1994; Tegmark,
Silk \& Blanchard 1995; Tegmark et al. 1996; Ostriker \& Gnedin 1996).
Since the ionization potential of hydrogen (13.6 eV) is 5--6 orders of
magnitude smaller than the nuclear energy release per baryon in a star
(6.6 MeV), only a negligible fraction of all the baryons need to be
processed through stars in order for the universe to get reionized.
An even smaller fraction is necessary if the progenitors of quasar
black holes form at the same time (Eisenstein \& Loeb 1995).

The reionization epoch begins as each ionizing source blows an
expanding Str\"omgren sphere around itself, and ends when these HII
regions overlap (Arons \& Wingert 1972).  However, long before the
Str\"omgren spheres overlap and breakthrough occurs, molecular
hydrogen gets universally dissociated by photons with energies in the
range $11$--$13.6$ eV, to which the neutral intergalactic gas is
transparent. As a result, molecular cooling is suppressed even inside
dense objects (Haiman, Rees, \& Loeb 1996).  Since efficient cooling
is required for stars to form through fragmentation of the metal--poor
primordial gas, the abundance of stars builds up to levels capable of
ionizing the universe only after massive objects ($\ga 10^8 M_\odot$)
form, inside which the high virial temperature of $\ga 10^4$K allows
fragmentation to occur through atomic line cooling.

Spectral observations of quasars at $z\la5$ show evidence for an early
reionization epoch. The lack of complete absorption shortward of the
\lya~resonance (the so--called Gunn--Peterson effect; Gunn \& Peterson
1965) implies that the universe is highly photoionized by $z\sim5$. The
source of the ionizing UV background is unknown, even at relatively low
redshifts. Several authors have noted that radiation from quasars alone
might be insufficient to reionize the universe (Shapiro \& Giroux 1987;
Miralda-Escud\'e \& Ostriker 1990), and recent observations of HeII
absorption favor a relatively soft spectrum of the ionizing
radiation, which is more typical of stars than of quasars (Vikhlinin 1995;
Haardt \& Madau 1996; but see Davidsen, Kriss \& Zheng 1996 for a different
view).

The expected reionization of the universe at $10\la z\la 10^2$ has a
number of observable consequences (see the early overview by Carr,
Bond \& Arnett 1984).  First, the resulting optical depth of the
universe to electron scattering, $\tau_{\rm es}$, damps the microwave
background anisotropies on scales $\la 10^\circ$ (Efstathiou \& Bond
1987; Kamionkowski, Spergel, \& Sugiyama 1994; Hu \& White 1996).  The
recently detected hint for a Doppler peak in ground--based microwave
anisotropy experiments can already be used to set the constraint
$\tau_{\rm es}\la 1$ (Scott, Silk \& White 1995; Bond 1995). Future
satellite experiments (such as MAP or PLANCK\footnote{See the
homepages for the MAP (http://map.gsfc.nasa.gov) and the PLANCK
(http://astro.estec.esa.nl/SA-general/Projects/Cobras/cobras.html)
experiments.})  will be able to probe or limit values of $\tau_{\rm
es}$ as small as $\sim0.1$.  Another consequence of the early epoch of
star formation is the metal enrichment of the intergalactic medium
(IGM).  Early star formation provides a likely explanation of the
metal abundance ($\sim 1\%Z_\odot$) which is detected at $z\sim3$ in
\lya~absorption systems with vastly different column densities
(Cowie~et~al.~1995; Tytler~et~al.~1995; Songaila \& Cowie 1996; Cowie
1996).  Although the level of enrichment in different systems varies
by an order of magnitude, the ubiquitous existence of heavy elements
in systems with low HI column densities ${\rm
N_{HI}\gsim10^{15}~cm^{-2}}$ is naturally explained by a
pre-enrichment phase (Songaila \& Cowie 1996). Finally, an old
population of pre-galactic stars would behave as collisionless matter
during galaxy formation and populate the diffuse halos of galaxies.
Such stars might explain some of the events observed by ongoing
microlensing searches in the halo of the Milky Way galaxy (see,
e.g.~Alcock et~al.~1996 and references therein; Flynn, Gould \&
Bahcall 1996) and could also be detected through their lensing effect
along lines of sight to distant quasars (Gould 1995).

In this paper, we quantify the observational signatures of the first
population of stars in CDM cosmologies.  We follow the formation of
star--forming regions out of primordial density fluctuations according
to the Press--Schechter theory, and calibrate the fraction of gas
which is converted into stars based on the inferred carbon abundance
in \lya~absorption systems.  Our semi--analytic approach incorporates
a variety of subtle gas dynamical processes, including the effect of
baryonic pressure on the abundance and sizes of virialized regions,
the significance of the initial mass function (IMF), the UV spectra and
metallicity yields of the stars, the negative feedback of
star--formation due to the photodissociation of molecular hydrogen,
the absorption of ionizing photons within the dense parent
star--forming clouds, and the temporal evolution of the ionization
fronts around star clusters.  Previous discussions in the literature
have discussed some, but not all, of these processes in full detail.

Since some of the details about star formation and radiative transfer
could {\it control} the solution to the reionization problem, a
semi--analytic approach appears to be best suited for sorting out the
significance of each of these ingredients under different model
assumptions.  Numerical simulations of the type described by Ostriker
\& Gnedin (1996) are expensive in computer time, and are still limited
by the approximate treatments of star formation and radiative
transfer.  In our calculation, we follow the evolution of the
metallicity of the intergalactic gas, the volume filling factor of
ionized regions, and the optical depths of the universe to electron
scattering and microlensing, over a large range of redshifts and mass
scales.  In contrast with numerical simulations, we are able to
calculate only the universal average of each quantity; however, the
versatility of our semi-analytic approach allows us to examine the
sensitivity of our results to many model parameters, such as the power
spectrum of primordial density fluctuations, the stellar IMF and
metallicity yields, the baryon density parameter $\Omega_{\rm b}$, the
escape fraction of ionizing photons from the parent star clusters, and
the presence of a negative feedback on star formation due to the
photo--dissociation of molecular hydrogen.

The paper is organized as follows. In \S~2 we provide a detailed
description of all the ingredients of our model.  In \S~3 we describe how
the various observable quantities (metallicity, optical depth to electron
scattering, and microlensing probability) are calculated from the model.  In
\S~4 we present our results and examine their sensitivity to model
parameters.  Finally, \S~5 summarizes the main conclusions and implications
of this work.

\section{Description of the Model}

In contrast with the sudden nature of standard recombination at
$z\approx 1400\pm 100$, reionization is a continuous phase transition
that lasts for a considerable fraction of its Hubble time. Initially,
most of the intergalactic medium (IGM) is neutral, and a number of
ionizing sources of different luminosities and lifetimes turn on.
Each of these sources creates an ionized bubble which expands into the
IGM at a rate dictated by the source luminosity and the background IGM
density.  Reionization is completed when these bubbles overlap to fill
the entire volume (see Arons \& Wingert 1972 for an early discussion
of this process in relation to quasars).

Energetically, it is sufficient to convert a small fraction of all the
baryons into stars ($\sim 10^{-3}$, see \S 2.3) in order to ionize the
IGM.  This implies that the universe could have been already ionized
when only a small fraction of its mass was incorporated into nonlinear
objects.  Therefore, at this stage, each of the rare star--forming
objects collapses in isolation and generates an HII bubble which
expands into a relatively smooth intergalactic medium.  Since the
lifetime of UV--bright stars is much shorter than the Hubble time at
$z\la 10^2$, the volume filling fraction of HII grows primarily due to
the rapid increase in the collapsed baryonic fraction with redshift.
The initial growth of this fraction is exponential for a Gaussian
random field of density fluctuations.  

The abundance and mass distribution of virialized dark matter halos is
described by the Press--Schechter theory (Press \& Schechter 1974).
Due to their finite initial pressure, the baryons are unable to
accrete onto shallow potential wells.  In our calculations, we take
into account the pressure of the baryons and employ spherically
symmetric simulations to calculate the abundance and structure of
baryonic objects with low masses (for a complete description of these
simulations see Haiman, Thoul \& Loeb 1996, hereafter HTL96).

As a result of the virialization process, the gas inside each collapsed
object settles into a smooth pressure--supported configuration.  Further
collapse and fragmentation of the gas into stars is possible only as a
result of additional cooling. Objects with masses near the Jeans mass
($\sim 10^5M_\odot$), are characterized by a virial temperature $\sim
100$--$300$K. At these temperatures, only cooling through the vibrational
and rotational transitions of molecular hydrogen (H$_2$) is efficient on a
dynamical time--scale in a gas of a primordial composition (HTL96).
However, the trace amounts of H$_2$ can be easily photodissociated by a
relatively weak flux of photons with energies between 11--13.6 eV, to which
the universe is transparent even before reionization (Haiman, Rees, \& Loeb
1997, hereafter HRL97).  The resulting negative feedback that suppresses
fragmentation into stars due to the photo--dissociation of $\rm{H_2}$
molecules, occurs shortly after the very first ionizing sources appear and
prevents further star formation until objects with virial temperatures
$\sim 10^4$K collapse, inside which atomic line cooling allows
fragmentation (HRL97).  Therefore, we assume in our standard model that
star formation occurs only inside objects with virial temperatures
$\ga10^4$K.  These objects are also immune to photo-ionization heating and
could therefore retain their gas even after the universe gets reionized.

The fraction of gas converted into pre--galactic stars determines the
average metallicity of the IGM.  While the local value of the metal
enrichment might vary from place to place due to inefficient mixing,
its line-of-sight average over cosmological scales should admit a
universal value.  This expectation is confirmed by the detection of a
roughly (to within an order of magnitude, cf. Rauch, Haehnelt \&
Steinmetz 1996; Hellsten et al. 1997) universal C/H ratio in
\lya~absorbers over a wide range of column densities at $z\sim 2$--$4$
(Cowie~et~al.~1995; Tytler~et~al.~1995; Songaila \& Cowie 1996, Cowie
1996).  The detected carbon might have been expelled out of an
abundant population of star forming regions with shallow potential
wells, through winds driven by supernovae or photoionization heating.

For a given mass function of stars, the inferred metallicity of the
IGM can be used to fix the overall fraction of gas which was converted
into stars.  In our model, we assume that this fraction is the same for
all gas clouds and arrange the total carbon output from all stars to
match the observed C/H ratio of $\sim 1\%$ solar at a redshift $z=3$.
We further postulate that star formation occurs in a starburst with a
universal Scalo (1986) initial mass function (IMF).  The consequences
of varying these model assumptions are described in \S 4.

The stellar IMF determines the composite spectrum of ionizing radiation
which emerges from the star clusters.  We follow the time-dependent
spectrum of a star of a given mass based on the Kurucz (1993) spectral
atlases and the evolution of the star on the H--R diagram as prescribed by
the tracks in Schaller~et~al.~(1992).  The ionizing photons which are
emitted from the stellar surfaces can still be consumed by recombinations
inside their parent gas cloud before they reach the IGM.  In modeling the
escape fraction of UV photons from a parent star--forming cloud, we adopt
the equilibrium density profile of gas inside each cloud according to our
spherically--symmetric simulations (HTL96).

Finally, the ionizing photons which escape each cloud generate an HII
bubble in the surrounding IGM. We use the time--dependent luminosity of
each star--forming region in calculating the propagation of this ionization
front.  Since the average separation between stellar sources at the time of
reionization is much shorter than the Hubble length, the propagation of
photons from a given source to its ionization front can be approximated as
instantaneous.  In our discussion, we ignore the possible contribution from
early quasars to the ionizing background (Eisenstein \& Loeb 1995). Our
account of the HII bubble dynamics around each time--dependent source
differs from several recent discussions which simply calculated the
ionization fraction of the universe based on the total energy input and the
chemistry of a uniform IGM (e.g.  Meiksin \& Madau 1993, Fukugita \&
Kawasaki 1994, Giroux and Shapiro 1996)

\subsection{The Number Density of Ionizing Sources}

The Press--Schechter theory relates the comoving number density,
$N_{\rm ps}(z,M)dM$, of collapsed objects with total masses between
$M$ and $M+dM$ at a redshift $z$, to the power spectrum of density
fluctuations $P(k)$ (see, e.g.~Padmanabhan~1993). In this paper we
adopt the BBKS power spectrum for CDM cosmologies
(Bardeen~et~al.~1986), which is parameterized by the primordial power
spectrum index $n$ and normalization $\sigma_{8h^{-1}}$.  We also
assume throughout the paper a cosmological density parameter of
$\Omega_0=\Omega_{\rm b}+\Omega_{\rm dm}=1$ and a Hubble constant of
$h\equiv H_0/100~{\rm km~s^{-1}~Mpc^{-1}}=0.5$.

In situations where gas pressure is negligible, the Press--Schechter theory
describes the behavior of the baryons as well. The comoving number density
of objects with a baryonic mass between $M_{\rm b}$ and $M_{\rm b}+dM_{\rm
b}$ is then given by $N_{\rm b}(z,M_{\rm b})=N_{\rm ps}(z,M_{\rm
b}\Omega_0/\Omega_{\rm b})$, where $\Omega_{\rm b}$ is the cosmological
density parameter of the baryons.  When incorporating the effect of gas
pressure, it is often assumed that no baryonic objects form below the Jeans
mass predicted by linear--theory $M_{\rm Jeans}\sim10^5{M_{\odot}}$, and so
$N_{\rm b}(z,M<M_{\rm Jeans})=0$. In reality, this assertion is inaccurate
as the gas pressure merely delays the collapse of low--mass baryonic
systems but does not prevent it.  Since the history of reionization is
sensitive to trace amounts of collapsed gas, it is important to include a
more accurate treatment of the gas dynamics in low mass objects which are
first to produce ionizing photons.  We incorporate the effect of gas
pressure into $N_{\rm b}(z,M_{\rm b})$ based on runs from our simulations
of spherically symmetric clouds (HTL96).  Our approach is the simplest
extension of the spherical top--hat collapse model which proved to be
successful when used to define the linear overdensity threshold for the
collapse of objects in the Press--Schechter mass function.  While the
top--hat model is adequate for defining the characteristic collapse time of
the dark matter, it must be modified for the baryons due to their pressure.

\begin{figure}[b]
\vspace{2.6cm}
\includegraphics{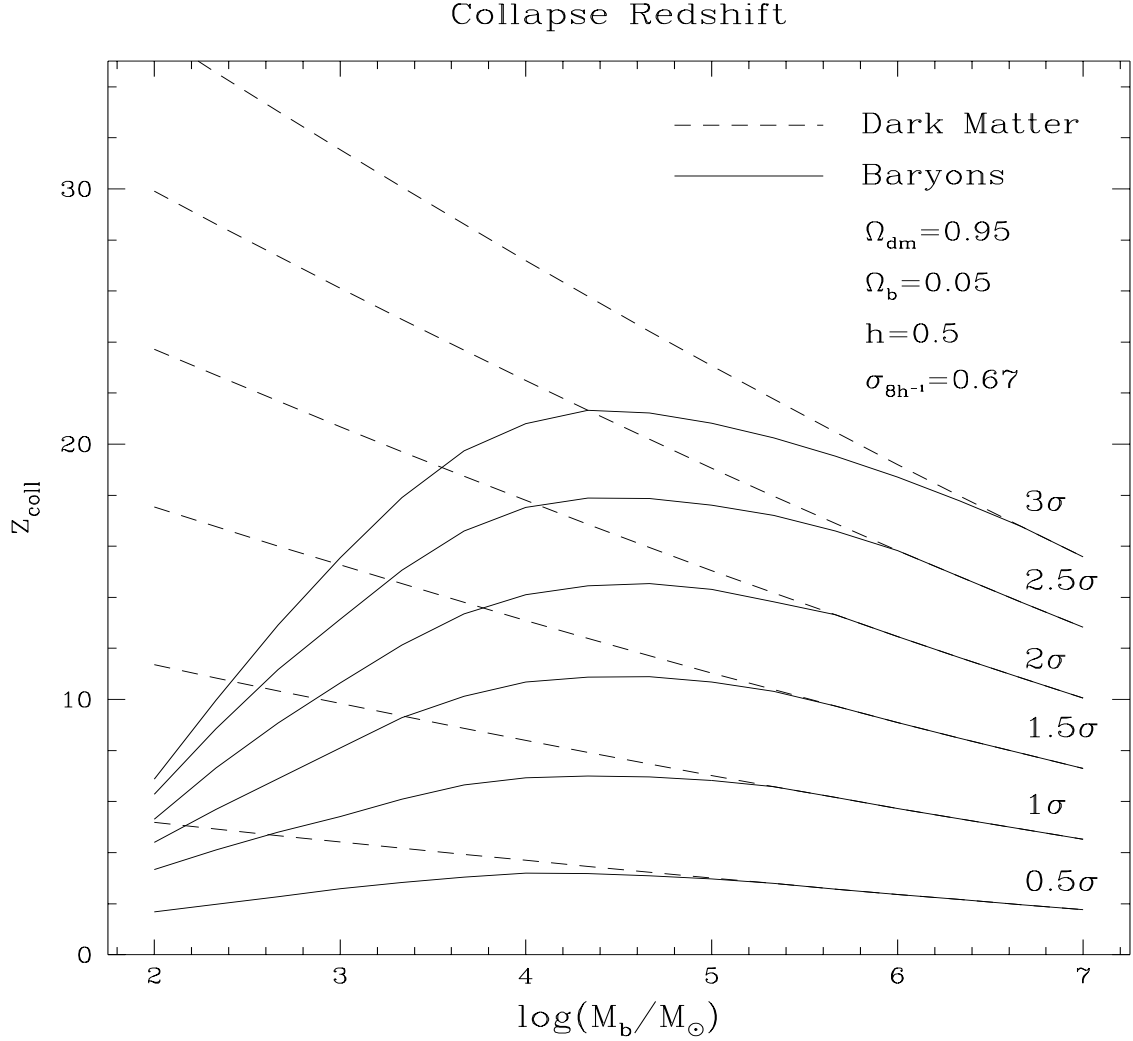}
\vspace*{4.5in}
\caption[Collapse Redshift] {\label{fig:zcoll} Collapse redshift for
cold dark matter (dashed lines) and baryons (solid lines) in spheres
of various masses and overdensities (the latter in units of
$\sigma(M)$ for a BBKS power spectrum). The collapse of the baryons is
delayed relative to the dark matter due to gas pressure.  The curves
were obtained by following the motion of the baryonic and dark matter
shells with a spherically symmetric, Lagrangian hydrodynamics code
(HTL96). The dashed lines agree with the collapse redshift for the
dark matter based on the linear extrapolation of the spherical
collapse model for pressureless matter, $z_{\rm
coll}=\nu\sigma/1.69-1$.}
\end{figure}

Our simulations start with a slightly overdense sphere shortly after
recombination.  The initial conditions are chosen to represent
spherically symmetric peaks of various widths and heights in the
linear overdensity field at $z=10^3$. If the mass of the sphere is
well above the Jeans mass, pressure can be ignored, and the baryonic
component expands, turns around, and collapses together with the dark
matter component.  However, for an object close or below the Jeans
mass, gas pressure delays the collapse of the baryons relative to the
dark matter. The smaller the mass of the system is, the longer the
delay gets.  We obtained the exact collapse redshifts by following the
motion of both the baryonic and the dark matter shells with our code.
Figure~\ref{fig:zcoll} shows the collapse redshift of both components
for spheres of different initial masses and overdensities.  This
figure demonstrates that objects with baryonic masses as low as ${\rm
\sim10^2M_{\odot}}$ (i.e., well below the Jeans mass predicted by
linear theory) are still expected to form before $z\sim20$.

\begin{figure}[b]
\vspace{2.6cm}
\includegraphics{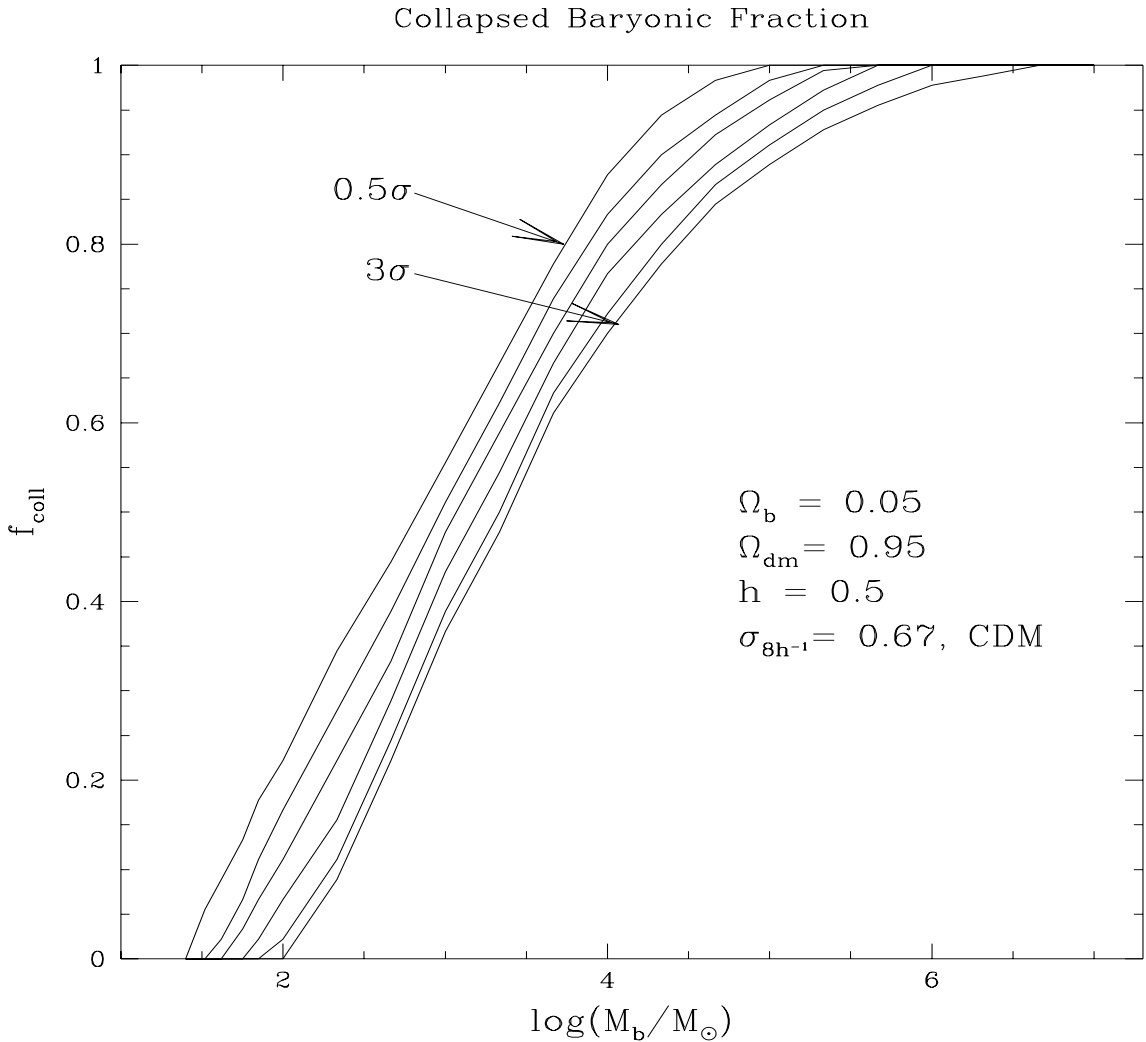}
\vspace*{4.5in}
\caption[Collapse Fraction] {\label{fig:fcoll} The fraction of baryons
that managed to collapse by the time the outermost shell of the dark
matter halo had collapsed.  The curves were obtained with the same
hydrodynamics code as in Figure~\ref{fig:zcoll}.  Six different curves
are shown for the six different peak heights, $0.5\sigma-3\sigma$, as
in Figure~\ref{fig:zcoll}.  These curves yield an ``equivalent
cosmological Jeans
mass'' of $10^3{M_{\odot}}$ (cf.~\S~2.1).}
\end{figure}

In order to incorporate the results shown in Figure~\ref{fig:zcoll} into
$N_{\rm b}(z,M_{\rm b})$, we determined in each run the fraction of the
baryonic mass which managed to collapse before the virialization of the
outermost dark matter shell.  Figure~\ref{fig:fcoll} shows this fraction as
a function of the total baryonic mass for various initial overdensities.
Based on the collapse fraction $f_{\rm coll}$, the number density of
objects is given by $N_{\rm b}(z,M_{\rm b})=N_{\rm ps}(z,M_{\rm
b}\Omega_0/f_{\rm coll}\Omega_{\rm b})$.

The total fraction $F_{\rm coll}$ of baryons contained within collapsed
objects at a redshift $z$ can then be expressed in terms of $f_{\rm coll}$
and $N_{\rm ps}$,
\beq 
F_{\rm coll}(z)=\frac{1}{\rho_0}\int_0^{\infty}dM 
f_{\rm coll}(z,M)M N_{\rm ps}(z,M),
\label{eq:fcoll}
\eeq 
where $\rho_0$ is the mean present-day density of the universe, and
$f_{\rm coll}$ is a function of the collapse redshift and mass of the
dark matter halo.  It is interesting to define an ``equivalent Jeans
mass'' by approximating $f_{\rm coll}$ as a step function with the
step located at the mass scale $M_{\rm crit}$ that provides the best
fit to the exact $F_{\rm coll}(z)$, obtained with the full $f_{\rm
coll}(z,M)$.  Using this definition, we obtain a critical mass $M_{\rm
crit}\approx10^{3}M_{\odot}$ before reionization, which is about 1--2
orders of magnitude smaller than the commonly adopted value for the
cosmological Jeans mass.

\subsection{Internal Structure of Ionizing Sources}

Our model assumes that a constant fraction $f_{\rm star}$ of the gas inside
each object turns into stars in an instantaneous starburst with a Scalo
(1986) IMF.  Since the IMF could be different at high redshifts than it is
observed locally, we also consider a tilted Scalo function which is
defined by adding a constant $\beta$ to the power law indices of the Scalo
IMF while keeping the IMF normalization fixed at 4${\rm M_{\odot}}$.  This
tilt leaves the total carbon yield almost unaffected but changes the net
fraction of gas which is converted into stars  (see discussion
below).

\begin{figure}[b]
\vspace{2.6cm}
\includegraphics{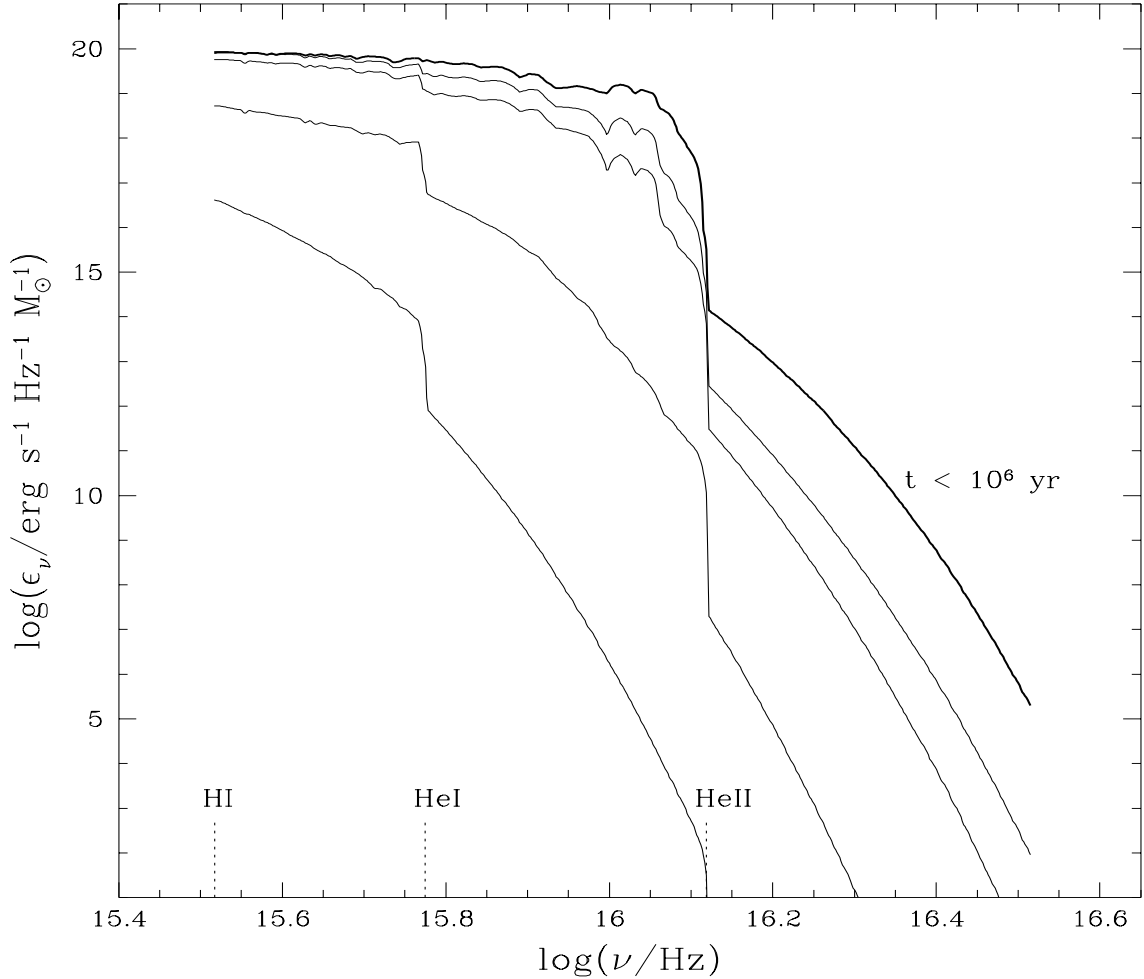}
\vspace*{4.5in}
\caption[Composite Spectrum] {\label{fig:spectrum} Composite
emissivity per solar mass for a star cluster with a Scalo IMF.  Each
star is assumed to evolve on the H--R diagram according to the
theoretical evolutionary tracks of Schaller~et~al.~(1992) and to emit
radiation according to the spectral atlas of Kurucz (1993).  The thick
line shows the steady emissivity when all the stars are still on the
main sequence ($\la 10^6$ years after the starburst).  The four solid
lines show the emissivity at four subsequent times (from top to
bottom): $t$=3, 4, 10, and 30 million years after the starburst.
Frequencies corresponding to the ionizations of hydrogen and helium
are marked -- note the sharp drop of the emissivity at the HeII edge.}
\end{figure}

For a star of a given mass, we find the time evolution of the
effective temperature ($T_{\rm eff}$) and surface gravity ($g_{\rm
surf}$) from the low metallicity ($Z=0.001$) evolutionary tracks in
Schaller~et~al.~1992.  For each pair of $T_{\rm eff}$ and $g_{\rm
surf}$ values, we then obtain the flux of the star $F(\nu)$ using the
spectral atlas of Kurucz (1993) at the lowest metallicity available,
$Z=10^{-5}$.  The resulting time--dependent composite emissivity
$\epsilon_{\nu}$ per solar mass is shown in Figure~\ref{fig:spectrum}.
The top curve shows the initial emissivity, which remains steady over
$\sim10^6$ years after the starburst.  Subsequently, the emissivity
drops as stars with progressively lower masses leave the main
sequence.

In this paper we ignore the emission of ionizing radiation from supernova
explosions.  Assuming that each star more massive than $8M_{\odot}$
explodes with a hydrodynamic energy output of $4\times10^{50}$ erg (Woosley
\& Weaver 1986), the supernova energy release per stellar mass, averaged
over a Scalo IMF, is $1.2\times10^{48}~{\rm erg}~M_{\odot}^{-1}$, i.e. only
$\sim 3\%$ of the energy released in stellar emission of ionizing photons
from the same IMF, $4.4\times10^{49}~{\rm erg}~M_{\odot}^{-1}$.

Figure~\ref{fig:flux} shows the expected evolution in the production rate
of ionizing photons for stars with various masses between 1 and
120$M_{\odot}$ (dotted curves) and for the composite flux of a Scalo IMF
(solid curve).  The ionizing flux starts to decline roughly two million
years after the starburst due to the limited lifetime of stars on the main
sequence.  The total number of ionizing photons which are released per
proton processed through a star is $\sim 1100$, about two order of
magnitudes lower than the corresponding value for a single O star.  The
dashed line shows the production rate of ionizing photons when the standard
IMF is tilted by a power--law index of $\beta$=1.69 in the direction of
introducing more low mass stars; this choice of $\beta$ is motivated by the
possibility that the halos of galaxies are composed of Massive Compact Halo
Objects (MACHOs).  The lack of massive stars in the tilted IMF reduces
significantly the number of ionizing photons to a total value of 0.66
ionizing photons per baryon processed through a star.

We assume that fragmentation converts a constant fraction of the gas into
stars at every point inside the cloud.  However, not all of the ionizing
photons which are emitted by these stars, escape the cloud and ionize the
IGM.  The escape probability $f_{\rm esc}$ depends on the density profiles
of both the gas and the stars inside the cloud. For both components, we
adopt a radial density profile that varies as $\propto 1/r^2$ and is
truncated at some outer radius, where the density jumps by a factor of
$\sim50$ from its background value.  This profile is in good agreement with
the numerical results described in HTL96, and is slightly shallower than
the dark matter profile ($\propto 1/r^{2.25}$) predicted by self--similar
solutions (Bertschinger 1985).  Under the assumption of a constant $f_{\rm
star}$ throughout the cloud, we calculate the equilibrium ionization
profile both inside and outside the cloud by solving the equations of
radiative transfer and ionization equilibrium, 
\beq
F_{\nu}(r)=\frac{1}{r^2}\int_{0}^{r} r^{\prime2}
e^{-\tau_{\nu}(r,r^{\prime})} j_{\nu}(r^{\prime})dr^{\prime},
\label{eq:transfer}
\eeq
\beq
\tau_{\nu}(r,r^{\prime})=\sigma_{\nu}
\left[N_{\rm H}(r)-N_{\rm H}(r^{\prime})\right],
\label{eq:tau}
\eeq
\beq
\alpha_{\rm B}(T)n_{\rm e}^2 = n_{\rm H} \int_{\nu_{\rm H}}^{\infty}
\frac{4\pi\sigma_{\nu}F_{\nu}}{h_{\rm P}\nu}d\nu,
\label{eq:ionbalance}
\eeq 
where $j_{\nu}$=$\epsilon_{\nu}f_{\rm star}\rho_{\rm b}(r)/(4\pi)$ is
the monochromatic emission coefficient in units of ${\rm erg}$ ${\rm
cm^{-3}}$ ${\rm s^{-1}}$ ${\rm ster^{-1}}$ ${\rm Hz^{-1}}$; $F_{\nu}$
is the flux in units of ${\rm cm^{-2}~s^{-1}}$;
$\tau_{\nu}(r,r^{\prime})$ is the optical depth between radii $r$ and
$r^{\prime}$; $\sigma_{\nu}$ is the ionization cross--section of
hydrogen in units of ${\rm cm^2}$; $N_{\rm H}(r)$=$\int_{o}^{r}n_{\rm
H}(r)dr$ is the column density of neutral hydrogen between the center
and a radius $r$; $\alpha_{\rm B}(T)$=$2.6\times10^{-13}~{\rm
cm^3~s^{-1}}$ is the case B recombination coefficient at
$T\approx10^4$K; $n_{\rm e}$ and $n_{\rm H}$ are the number densities
of electrons and neutral atoms in ${\rm cm^{-3}}$; $\rho_{\rm b}(r)$
is the mass density profile of the gas; $\nu_{\rm H}$ is the ionization
threshold of hydrogen, and $h_{\rm P}$ is Planck's constant.

\begin{figure}[b]
\vspace{2.6cm}
\includegraphics{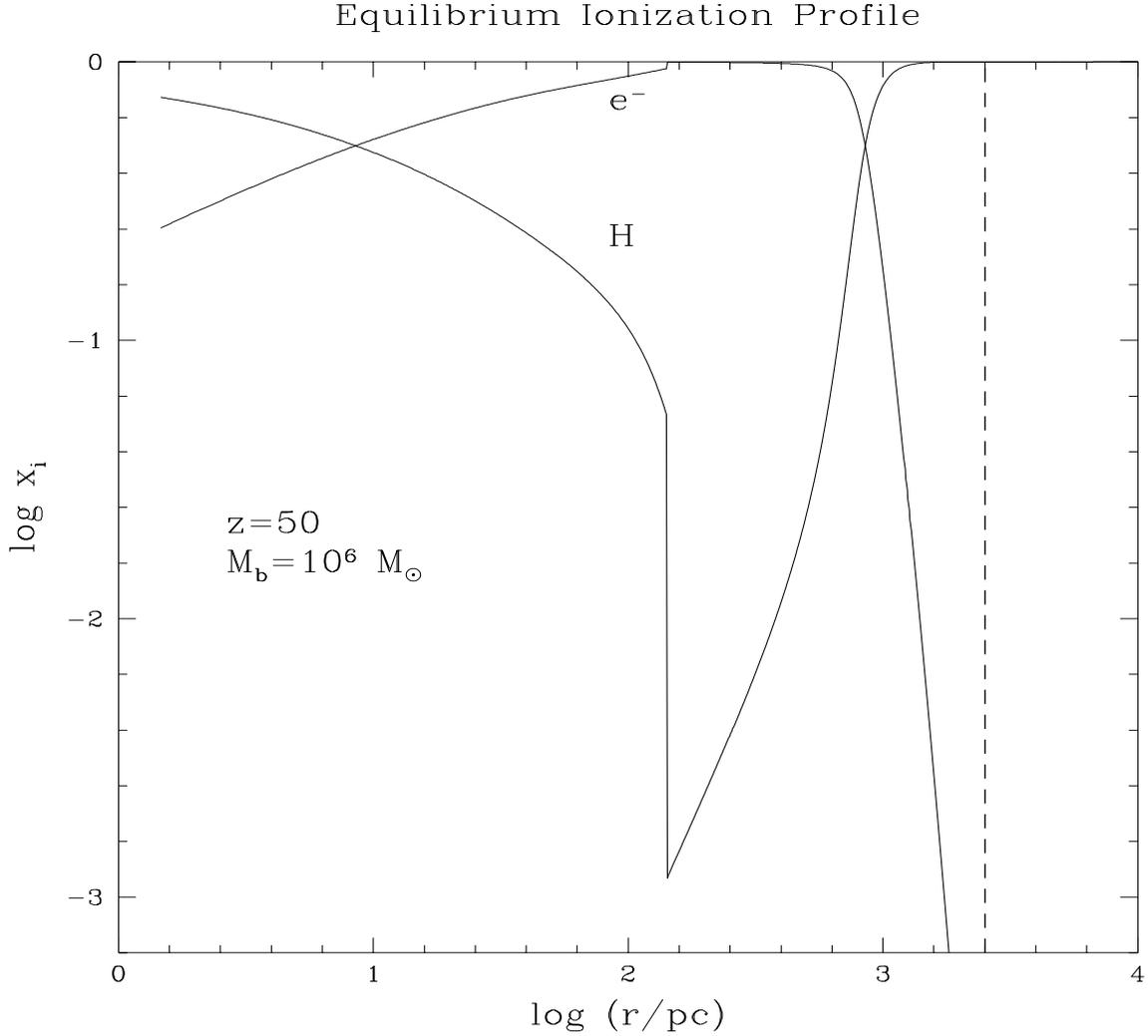}
\vspace*{4.5in}
\caption[Equilibrium ionization Profile] {\label{fig:ionprof} Example
of an equilibrium ionization profile inside and around a spherically
symmetric object: the fractions $x_{\rm e}=n_{\rm e}/(n_{\rm e}+n_{\rm
H})$ and $x_{\rm H}=n_{\rm H}/(n_{\rm e}+n_{\rm H})$ are shown as a
function of radius. We assume that the object has a baryonic mass of
$10^6{M_{\odot}}$ and a density profile $\propto1/r^2$, with a factor
of 50 drop in the density at the edge of the object ($r\sim150$pc),
outside of which the density is uniform and equal to the background
value at $z$=50.  A fraction $f_{\rm star}=13\%$ of the
baryons in the object are converted into stars with the emission
spectrum shown in Figure~\ref{fig:spectrum}.  The vertical dashed line
shows the radius of the Str\"omgren sphere when recombinations in the
object are ignored; this can be used to infer the escape fraction
of ionizing photons of 3.3\%.}
\end{figure}

Figure~\ref{fig:ionprof} shows an example of the resulting ionization
profile for an object with a baryonic mass of $10^6{M_{\odot}}$ at
$z=50$.  The size of this object is $\sim 150$ pc, and the vertical
dashed line at $\sim2.5$ kpc shows the size that the associated
Str\"omgren sphere would have had if the recombinations inside the object
were ignored and only the recombinations in the background gas were
taken into account.  The recombinations inside the object reduce the
size of the Str\"omgren sphere to $\sim 0.8$kpc, from where we infer
that only a fraction $f_{\rm esc}\approx (0.8/2.5)^3=3.3\%$ of the
ionizing photons escaped from the object.

\begin{figure}[b]
\vspace{2.6cm}
\includegraphics{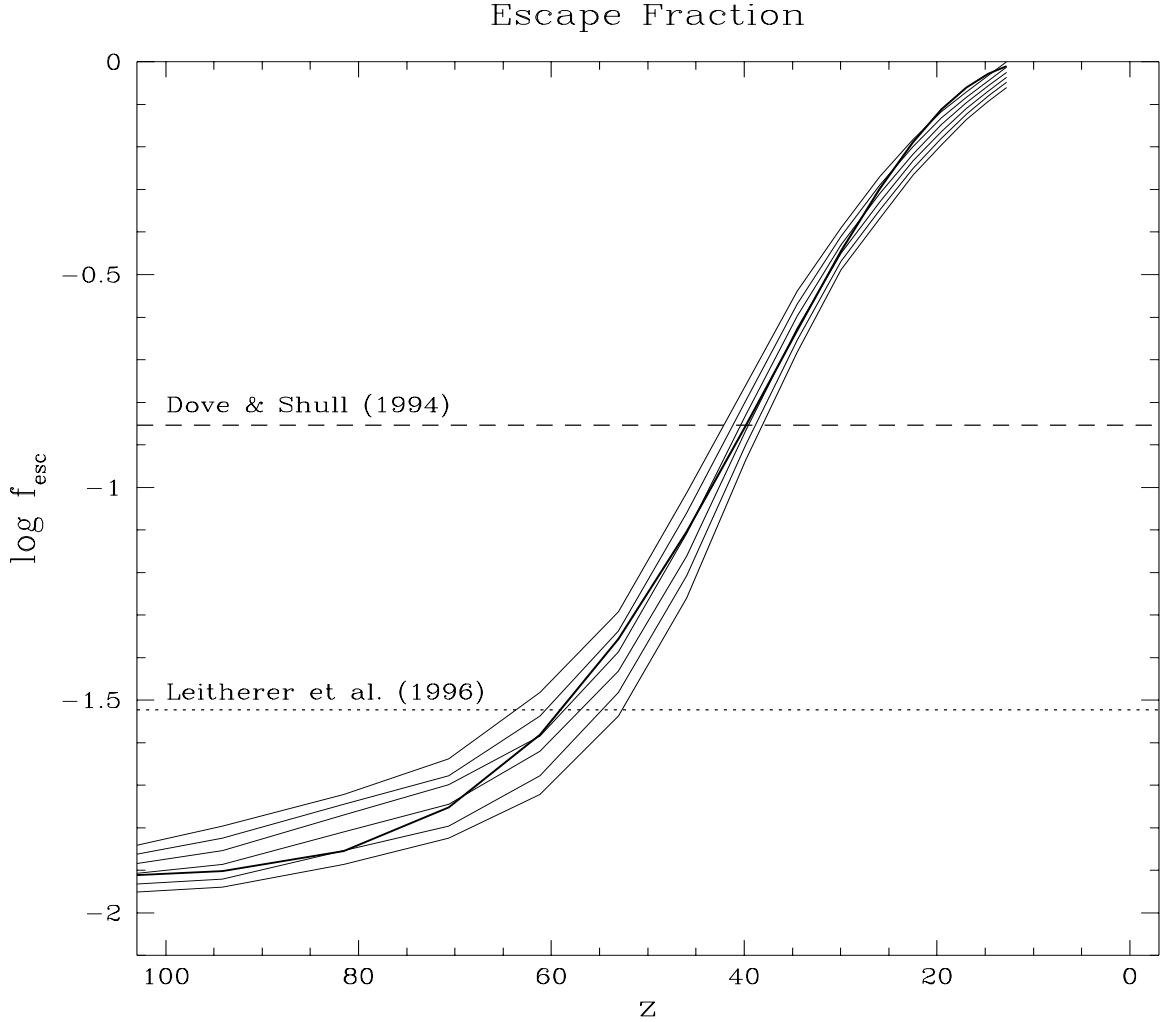}
\vspace*{4.5in}
\caption[Escape Fraction] {\label{fig:escape} Escape fraction of
ionizing photons from objects of different masses at various
redshifts.  The curves from bottom to top correspond to baryonic
masses ${10^{4,5,6,7,8,9}~M_{\odot}}$, and are obtained from
equilibrium ionization profiles similar to Figure~\ref{fig:ionprof}.
The boldface solid line shows the fitting formula $\log f_{\rm
esc}=1.92 (\exp[-(z-10)^2/1510]-1)$ for $z>10$. For comparison, the
dashed line shows the theoretical estimate of the escape fraction from
the disk of the Milky Way galaxy (Dove \& Shull 1994), and the dotted
line shows the observed value of the escape fraction in four nearby
starburst galaxies (Leitherer~et~al.~1996).}
\end{figure}

We have computed similar equilibrium ionization profiles for objects
of different masses and redshifts, and in each case we obtained the
escape fraction $f_{\rm esc}$.  The results are shown in
Figure~\ref{fig:escape} for objects with masses between
$10^4M_{\odot}$ (top curve) and $10^9M_{\odot}$ (bottom curve).  This
figure demonstrates that the escape fraction varies only slightly with
mass but depends strongly on redshift.  At redshifts below $z=10$ the
escape fraction is close to unity; at higher redshifts one can
derive a useful fitting
formula, accurate to within $\sim$ 10\%,
\beq
\log f_{\rm esc}=
\left\{\matrix{
1.92 \{\exp[-(z-10)^2/1510]-1\}\hfill&({\rm for}~z>10)\hfill\cr
0\hfill&({\rm for}~z\le 10)\hfill.\cr}\right.
\label{eq:escape}
\eeq

Although the escape fraction of the ionizing photons is a crucial
ingredient in all of our reionization models, little is known about its
value in nearby star forming regions.  For reference, it is possible to
compare our estimates with the value of 14\% obtained by Dove \& Shull
(1994) in their theoretical modeling of HII regions around OB associations
inside the exponential disk of the Milky Way galaxy (short dashed line in
Fig.~\ref{fig:escape}).  Another possible comparison is with the value of
3\% reported by Leitherer~et~al.~(1995) from observations of four nearby
starburst galaxies (long dashed line). On the one hand, our approach
overestimes the escape fraction because it ignores small scale clumpiness
of the gas. On the other hand, it underestimates the escape fraction,
because the system might lose its gas due to winds driven by supernova
explosions or photoionization heating.  The dependence of our results on
changes in the adopted values of the escape probability is discussed in
\S4.

\subsection{Propagation of the Ionization Fronts}

The propagation of ionization fronts in an expanding medium was
considered by Shapiro \& Giroux (1987) in the context of reionization
of the universe by quasars.  The temporal evolution of the radius of
the ionization front $r_{\rm i}$ in physical (i.e. not comoving)
coordinates is determined by the equation
\beq n_{\rm H}\left(\frac{dr_{\rm i}}{dt}-H(t)r_{\rm i}\right)=
\frac{1}{4\pi r_{\rm i}^2}\left(\frac{dN_{\gamma}}{dt}(t)-
\frac{4}{3}\pi r_{\rm i}^3 \alpha_{\rm B}n_{\rm H}^2 \right),
\label{eq:front}
\eeq 
where $H(t)$ is the Hubble constant, ${dN_{\gamma}/{dt}}(t)$ is the
emission rate of ionizing photons by the source, and $n_{\rm H}$ is
the background number density of neutral hydrogen in the IGM.
Equation~(\ref{eq:front}) simply expresses the fact that during the
expansion of the ionization front, the ionization rate of fresh atoms
is equal to the difference between the emission rate of photons from
the source and the total recombination rate in the already ionized
region.  In the case of a steady source ($dN_{\gamma}/dt$=const),
equation~(\ref{eq:front}) can be solved analytically (Shapiro \&
Giroux 1987) for $r_{\rm i}(t)$; with a time-dependent source, we
substitute ${dN_{\gamma}}/{dt}(t)$ and obtain $r_{\rm i}(t)$
numerically.  In our model,
\beq
\frac{dN_{\gamma}}{dt}(t)= f_{\rm esc} f_{\rm star} M_{\rm b}
\int_{\nu_{\rm H}}^{\infty} d\nu \frac{\epsilon_{\nu}}{h_{\rm P}\nu},
\label{eq:source}
\eeq 
where $M_{\rm b}$ is baryonic the mass of the source, $f_{\rm esc}$ is
the escape fraction of photons from the source, $f_{\rm star}$ is the
fraction of gas converted to stars, and $\epsilon_{\nu}$ is the
composite stellar emissivity per unit mass.

\begin{figure}[b]
\vspace{2.6cm}
\includegraphics{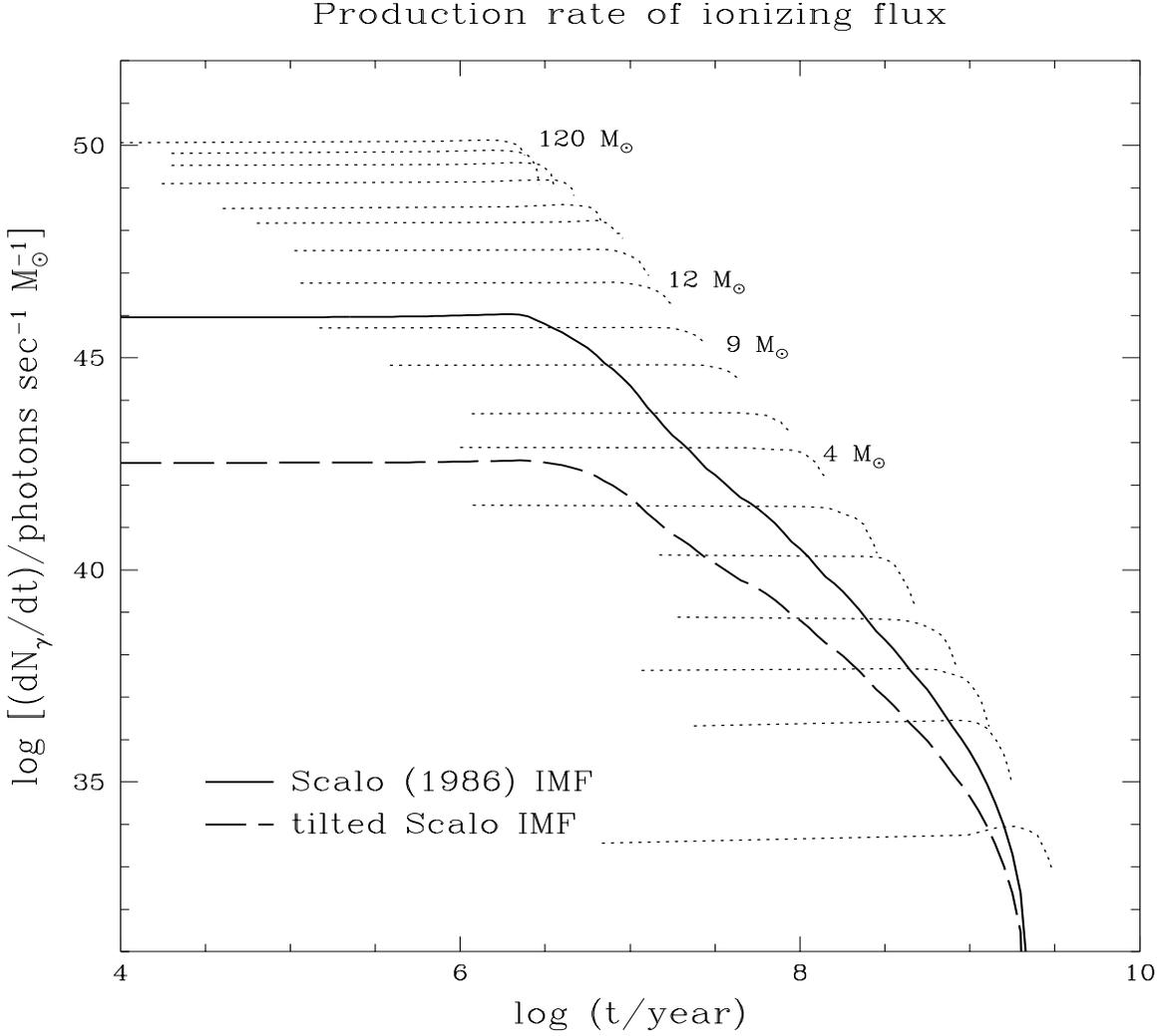}
\vspace*{4.5in}
\caption[Composite Photon Flux] {\label{fig:flux} The boldface solid
line shows the number of ionizing photons emitted per second per solar
mass (without the factors $f_{\rm esc}$ and $f_{\rm star}$, see
\S~2.3) for the composite emissivity shown in
Figure~\ref{fig:spectrum} and a Scalo IMF.  The boldface dashed line
shows the rate for a tilted Scalo IMF, where most of the stars have 
low masses.  The dotted lines show the separate contributions from stars
with masses (top to bottom) 120, 85, 60, 40, 25, 20, 15, 12, 9, 7, 5,
4, 3, 2.5, 2, 1.7, 1.5, and 1.25 $M_{\odot}$, in units of
photons/sec.}
\end{figure}

\begin{figure}[b]
\vspace{2.6cm}
\includegraphics{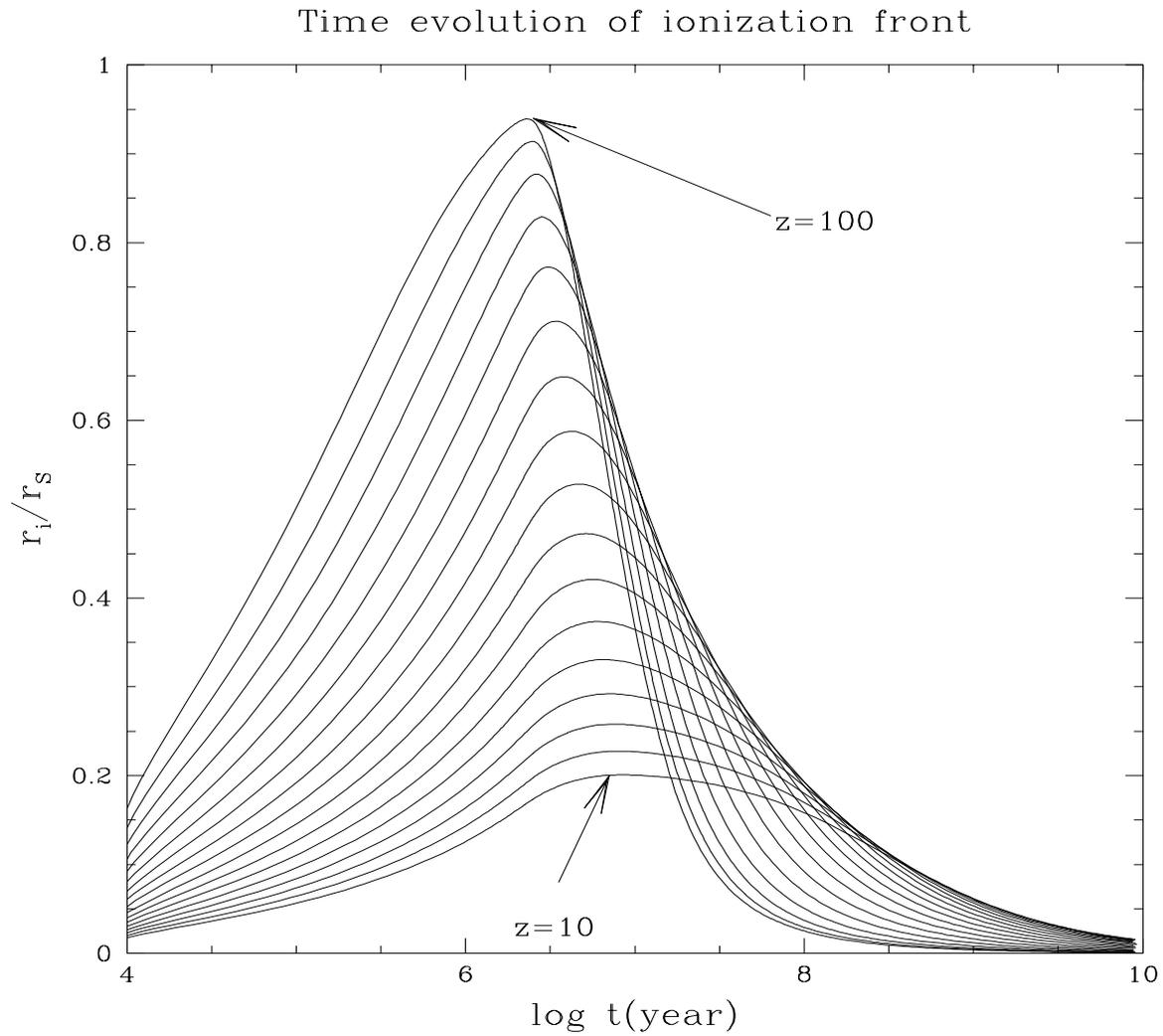}
\vspace*{4.5in}
\caption[Ionization Fronts] {\label{fig:fronts} Radius of the
ionization front in units of the time-varying Str\"omgren radius
(cf.~eq.~\ref{eq:strom}), for sources that turn on between redshifts
$z$=10 and $z$=100, and have the photon production rate shown by the
solid line in Figure~\ref{fig:flux}.  The turn-on redshifts of the 17
curves are equally spaced in log $z$.}
\end{figure}

It is instructive to examine the evolution of the ionization front in
units of the time--dependent Str\"omgren radius of a steady source
at the initial luminosity, 
\beq 
r_{\rm s}=\left[\frac{3}
{4\pi\alpha_{\rm B}n_{\rm H}^2(t)}{dN_{\gamma}\over dt}(t=0)\right]^{1/3}.
\label{eq:strom}
\eeq
Figure~\ref{fig:fronts} illustrates the evolution of the ionization front
in units of the Str\"omgren radius for sources that turn on at different
redshifts between 10--100, and have the ionizing photon production rate
shown in Figure~\ref{fig:flux}.  Two interesting effects are apparent:

\begin{itemize}
\item[1.] The ionization front reverses its direction of propagation 
when the flux from the source starts to decline
(cf. Fig.~\ref{fig:flux}). The decrease in the volume of the HII region
can be substantial at high redshifts, when the recombination time in the
IGM is short.

\item[2.] At redshifts $z\la 10^2$, the ionization front does not have 
sufficient time to expand to its full Str\"omgren radius, as pointed
out by Shapiro and Giroux (1987). This effect is less important at high
redshifts, because the recombination time in the IGM is then shorter and 
the Str\"omgren radius is therefore closer to the source.
\end{itemize}

Note that our calculation assumes that the ionization front propagates
out from a virialized nonlinear object into a uniform IGM.  Numerical
simulations (such as described in Gnedin \& Ostriker 1996) are
necessary in order to explicitly incorporate the mildly nonlinear IGM
clumpiness in this calculation.

\subsection{Carbon Production and Star Formation Efficiency}

Because we presently lack a standard theory of star formation, our
models (as well as other studies, such as Fukugita \& Kawasaki 1994,
and Madau \& Shull 1996) simply assume that a constant fraction
$f_{\rm star}$ of the gas is transformed into stars.  Our approach is
to calibrate $f_{\rm star}$ based on the inferred carbon abundance in
the IGM from Keck observations of the \lya~forest (Cowie~et~al.~1995;
Tytler~et~al.~1995; Songaila \& Cowie 1996, Cowie 1996).  The low
column density absorbers in the \lya~forest extend over large
transverse scales of $\ga 0.1$--$1$ Mpc (Bechtold et al. 1994; Dinshaw
et al. 1994, 1995; Fang et al. 1995; Smette et al. 1995) and can be
interpreted as mildly overdense regions in a photoionized IGM
(Hernquist et al. 1996, Miralda-Escud\'e et al.  1995).  We assume
that the apparently universal metallicity of these systems resulted
from metal enrichment by many low--mass ($\sim 10^{8-9}M_\odot$) star
clusters inside of them.  In our standard model we adopt the most
conservative approach which assumes that all the metals produced by
the stars are mixed efficiently with the IGM; a lower mixing
efficiency would imply a higher value for $f_{\rm star}$ and an even
earlier reionization epoch for the same value of IGM metallicity.  The
required mixing of the metal rich gas with the IGM could have resulted
from supernova driven winds, since the velocity dispersion of the
systems which are responsible for reionization in our models is only
$\sim 10~{\rm km~s^{-1}}$.  However, even if some metallicity
gradients remained because of incomplete mixing, the line-of-sight
integral inherent to observations of the \lya~forest systems tends to
average over these variations and yield a roughly universal
metallicity value throughout the universe.

We assume that the same population of stars that produced the heavy
elements detected in the \lya~absorbers is responsible for the
reionization of the universe.  In order to fix $f_{\rm star}$, we need
to specify the amount of heavy elements produced by a population of
stars with the Scalo IMF.  Since the Keck observations detected
specifically CIV absorption lines, we focus our attention on stellar
yields of $^{12}$C.  The carbon yields of high mass stars are well
established in the literature.  Weaver \& Woosley (1993) have
calculated the yields for stars with masses between
11--40${M_{\odot}}$, and their numbers are in good agreement with the
yields reported by Maeder (1992) for stars with 9--120${M_{\odot}}$.
For low and intermediate mass stars with 1--8${M_{\odot}}$, the yields
were tabulated by Renzini \& Voli (1981), and more recently by van den
Hoek \& Groenwegen (1996).  However, since these stars do not explode
as supernovae, there is a significant uncertainty about the amount of
carbon that actually gets ejected from these stars to the surrounding
medium (e.g.~through thermal pulsations or winds).  This uncertainty
is particularly large for stars which are more massive than
$\sim3{M_{\odot}}$, as these stars are known to undergo some amount of
``hot bottom burning'', i.e. burning of carbon into nitrogen at the
base of their convective envelopes. The actual amount of hot bottom
burning depends sensitively on the convective mixing length, and is
highly uncertain.

\begin{figure}[b]
\vspace{2.6cm}
\includegraphics{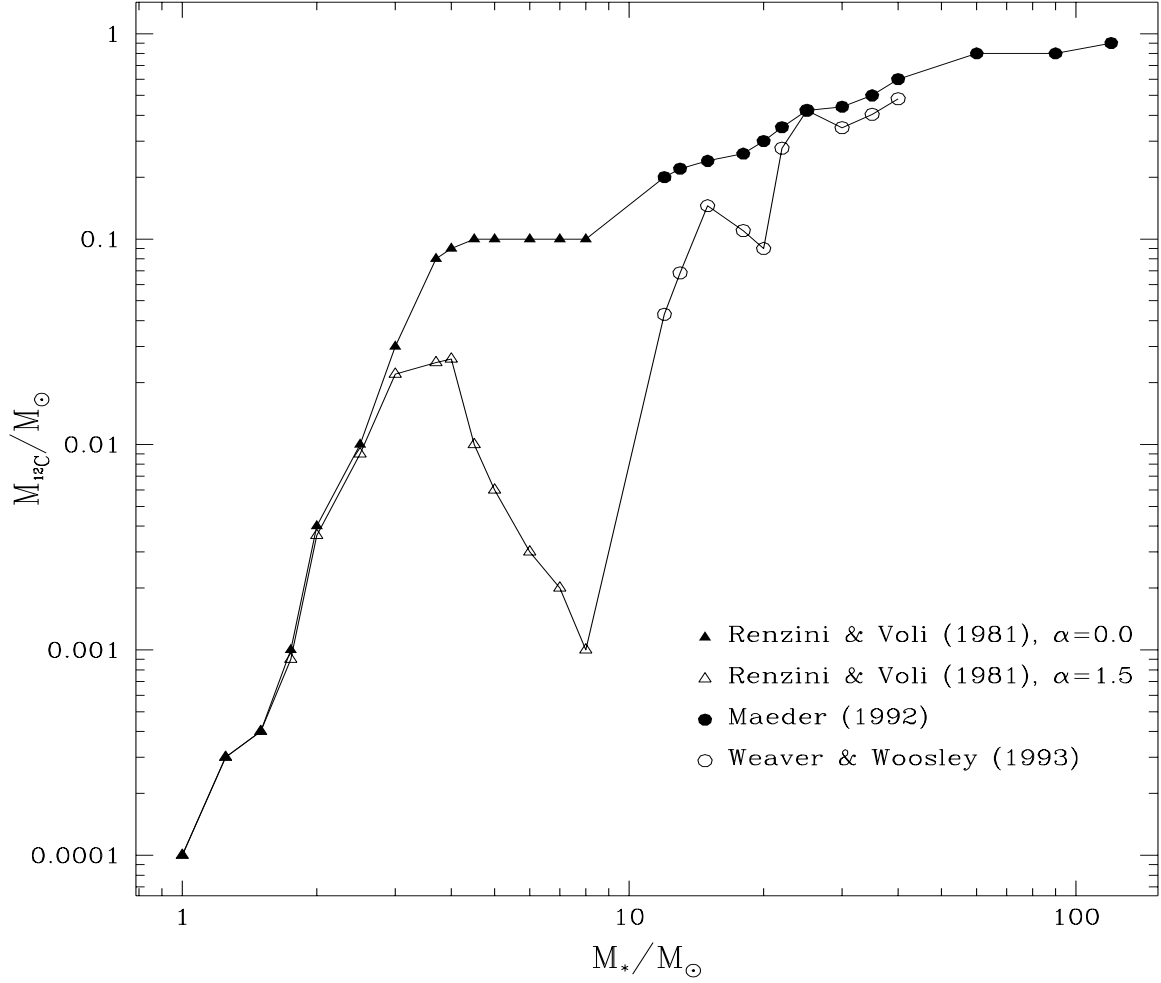}
\vspace*{4.5in}
\caption[carbon Yields] {\label{fig:yields} Carbon yields (${\rm
M_{^{12}C}}$ in solar masses) from stars of different masses.  The
large allowed range in the yields between 3--8 ${\rm M_{\rm \odot}}$
in Renzini \& Voli (1981) is due to the uncertainty in the extent of
hot bottom burning; $\alpha=1.5$ and 0 correspond to efficient, or no
hot bottom burning, respectively.}
\end{figure}

\begin{figure}[b]
\vspace{2.6cm}
\includegraphics{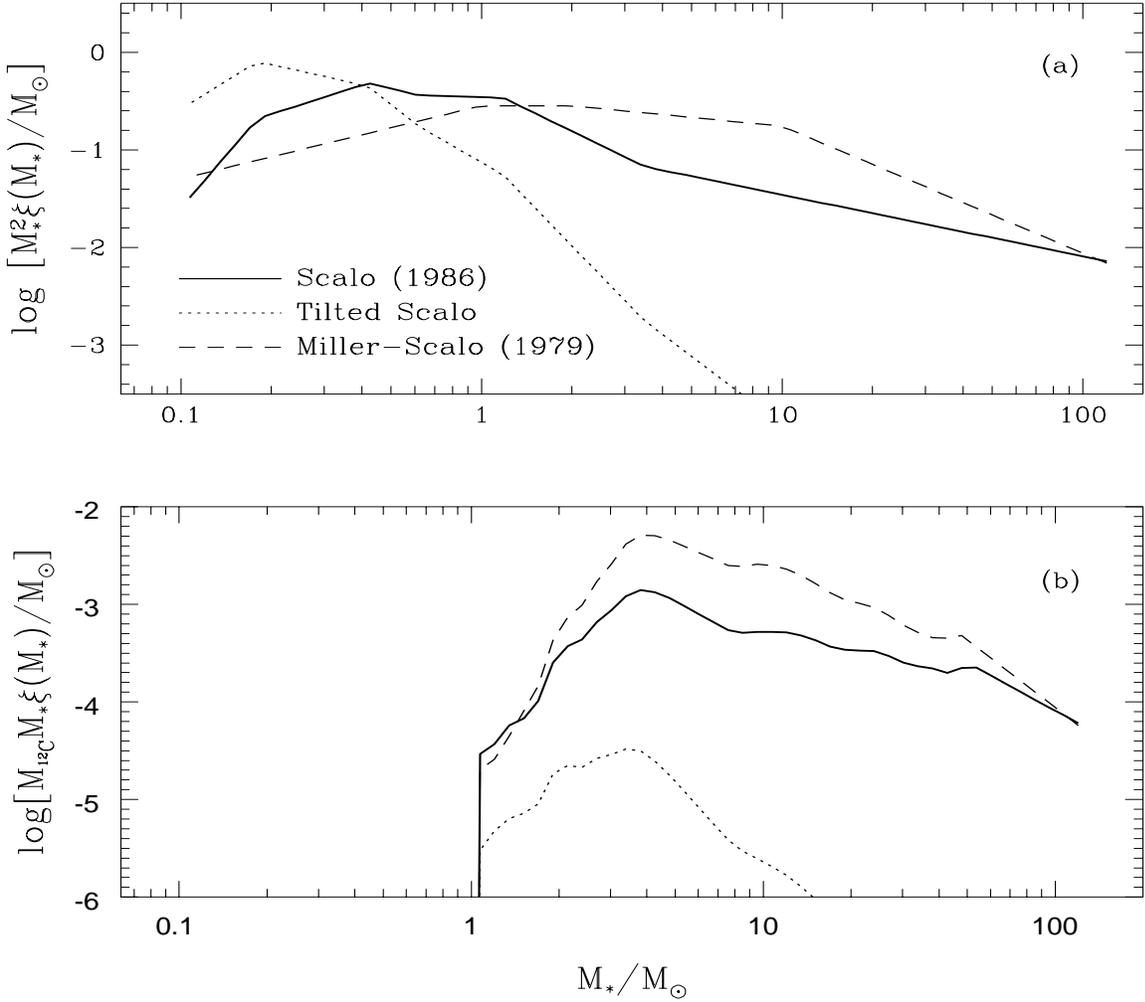}
\vspace*{4.5in}
\caption[IMF] {\label{fig:imf} Upper panel: three different initial
mass functions (IMFs). The Scalo (1986) and Miller-Scalo (1979) IMFs
are two different parameterizations of the locally observed stellar
population, and the tilted Scalo IMF is obtained by adding a constant
$\beta$=1.69 to the power law index in each segment of the Scalo IMF,
while keeping the value of the IMF fixed at $4{M_{\odot}}$.  All three
IMFs are normalized to an integral of $1{M_{\odot}}$. Lower panel:
carbon yields from stars of different masses (${\rm M_{^{12}C}}$, in
solar masses), folded in with the three different IMFs.  In all cases,
most of the carbon is produced by $\sim3$--$6{M_{\odot}}$ stars.}
\end{figure}

Figure~\ref{fig:yields} shows the carbon yields as a function of
stellar mass.  The large range of yields allowed for 3--8${M_{\odot}}$
stars is due to the unknown efficiency of hot bottom burning.
Additional uncertainty in the total carbon yield results from the
unknown stellar IMF; Figure~\ref{fig:imf} shows the {\it maximum}
allowed carbon yields folded in with a Scalo (1986), Miller--Scalo
(1979), and tilted Scalo IMFs. In our standard model we adopt the
first choice.  The sensitivity of our results to variations in these
uncertain parameters will be examined in \S~4.

Figures~\ref{fig:yields} and \ref{fig:imf} show that most of the
carbon is produced by 3--6${M_{\odot}}$ stars, and therefore the total
amount of carbon output by all stars is subject to the uncertainties
about hot bottom burning.  In order to get an average carbon
enrichment equivalent to 1\%$Z_{\odot}$, as measured in the
\lya~forest, it is necessary that a nominal 2.8\% of all the baryons
be converted into stars with a Scalo IMF (or a fraction 1.2\% for a
Miller--Scalo IMF).  The corresponding required value of the
star--formation efficiency for the Scalo IMF is $f_{\rm
star}=13\%$. Here, a factor of $\sim2.3$ increase is from the average
time required to produce carbon inside the stars; i.e. only a fraction
1/2.3 of the total stellar carbon is produced by $z=3$; another factor
of 2 increase is from the value of the collapsed fraction, $F_{\rm
coll}=50\%$ at $z=3$.  Note that $f_{\rm star}$ must be increased
further if any significant amount of hot bottom burning occurs in
stars with 3--8${M_{\odot}}$.

Finally, it is interesting to note that most of the carbon enrichment
is due to stars with 3--6${M_{\odot}}$, while most of the ionizing
photons are produced by stars with 10--12${M_{\odot}}$.  Because of
this fact, the carbon enrichment is delayed considerably relative to
the photon production; 90\% of ionizing photons are emitted already
within $\sim 5\times10^7$ years following the starburst, while it
takes $\sim5\times10^8$ years to eject 90\% of the total carbon
output.

\section{Consequences of the Model}

Next, we use the model described in the previous section to calculate
the redshift evolution of various integral quantities, i.e.~the
fraction of baryons contained in collapsed objects, the average
metallicity of the IGM, the volume filling factor of ionized regions,
the microlensing probability, and the optical depth to electron
scattering which translates to a damping factor for the cosmic
microwave background (CMB) anisotropies. Below we discuss each of
these quantities separately.

\vspace{0.5cm}
\noindent 
{\it The Collapsed Fraction}

The fraction of baryons contained in collapsed objects is given in
equation~(\ref{eq:fcoll}).  In our model, a fixed universal fraction
$f_{\rm star}$ of this material is converted into stars.  However, as
argued in \S 2, fragmentation of the collapsed clouds into stars is
initially inhibited by the photo--dissociation of ${\rm H_2}$ (cf. HRL97).
Shortly after the very first ionizing sources turn on, ${\rm H_2}$ cooling
is suppressed even in dense environments.  Star formation resumes when
clouds with a virial temperature $\ga 10^4$K, or a total mass $\ga
10^{8}{M_{\odot}}[(1+z)/10]^{-3/2}$ collapse, inside which atomic line
cooling becomes efficient and triggers fragmentation.  We incorporate this
pause in the star-formation history into our model by setting $f_{\rm
star}=0$ in equation~(\ref{eq:fcoll}) for clouds with masses below
$10^{8}{M_{\odot}}[(1+z)/10]^{-3/2}$.  Note, however, that star formation
is suppressed by a factor of $\sim[1-F_{\rm HII}(z)]$ even in the absence
of ${\rm H_2}$-feedback; the collapse of clouds in a hot ($\sim 10^4~{\rm
K}$) photoionized gas is delayed until dark-matter potential wells with
sufficient depth form (Couchman \& Rees 1986).  We therefore multiply
$f_{\rm star}$ by $[1-F_{\rm HII}(z)]$ in one of the variants on our
standard model where ${\rm H_2}$-feedback is ignored.

\noindent
{\it Metallicity}

We obtain the average metallicity of the IGM from the carbon yields as
described in \S~2.4.  Our standard model assumes that each star deposits
all of its carbon yield in the background gas at the end of its life on the
main sequence, and that the metal enriched gas is fully mixed with the
IGM\footnote{A more refined modeling might treat the dependence of the
``mixing efficiency'' with the IGM on the sizes of the aggregates in which
stars form (Rees 1996).}. We define $f_{\rm carb}(\Delta t)$ to be the
fraction of the total carbon yield which gets deposited after a time
$\Delta t$ following the starburst.  As a result of this deposition, the
IGM obtains a carbon metallicity
\beq
Z(z)=0.01Z_{\odot}\times\left(\frac{f_{\rm star}}{0.028}\right)
\int_{\infty}^{z}dz^{\prime}\frac{dF_{\rm coll}}{dz}
(z^{\prime})f_{\rm carb}\left[t(z)-t(z^{\prime})\right].
\label{eq:metals}
\eeq 
The value $f_{\rm star}=0.028$ would result in the required
metallicity $0.01Z_{\odot}$ if the collapsed fraction $F_{\rm coll}$
was unity and there was no time--delay in the carbon prodcution
(assuming the maximum possible carbon yields with a Scalo IMF,
cf. \S~2.4). $Z(z)$ is proportional to $F_{\rm coll}(z)$ only at low
redshifts when the finite time needed to deposit the carbon yields is
much shorter than the Hubble time.

\noindent
{\it HII Filling Factor}

Our model can be used to determine the fraction of the volume of the
universe in which hydrogen is ionized as a function of redshift. We
define $r_{\rm i}(z_{\rm on},z,M)$ to be the time dependent radius of
the spherical ionization front around a source of mass $M$ which turns
on at a redshift $z_{\rm on}$.  Equation~(\ref{eq:front}) can then be
used to determine $r_{\rm i}(z_{\rm on},z,M)$ for a range of $z$ and
$M$ values.  Since the escape fraction of ionizing photons
$f_{\rm esc}$ is, to leading order, independent of the mass of the
system, the volume ionized by a given system scales linearly with its mass,
namely $V_{\rm HII}(z_{\rm on},z,M)=(4\pi r_{\rm i}^3/3)
= M\times {\tilde V}_{\rm HII}(z_{\rm on},z)$, where
${\tilde V}_{\rm HII}(z_{\rm on},z)$ is the ionized volume per unit mass.
We can therefore evaluate the HII filling factor $F_{\rm HII}(z)$
from the collapsed gas fraction 
\beq
F_{\rm HII}(z)=\rho_{\rm b}(z)
\int_{\infty}^{z}dz^{\prime}\frac{dF_{\rm coll}}{dz}
(z^{\prime})  {\tilde V}_{\rm HII}(z^{\prime},z) ,
\label{eq:filling}
\eeq 
where $\rho_{\rm b}=\Omega_{\rm b}\rho_{\rm crit}$ is the average
baryonic density and $\rho_{\rm crit}$ is the critical density of the
universe. Note that this result depends only the total fraction 
of baryons in nonlinear objects and not on the merger
history of these objects.

\noindent
{\it Microlensing Optical Depth}

The probability for having a microlensing event along the line-of-sight to
a point source located at a redshift $z_{\rm S}$ in an $\Omega_0=1$
universe, is given by,
\beq
\tau_{\rm lens}(z_{\rm S})=\frac{\Omega_{\rm b}\rho_{\rm crit}c}{H_0}
\int_o^{z_{\rm S}} dz_{\rm L}
\sqrt{1+z_{\rm L}} f_{\rm star} F_{\rm coll}(z_{\rm L}) 
\sigma_{\rm lens}(z_{\rm L}),
\label{eq:taulens}
\eeq 
where $c$ is the speed of light, and the cross--section per unit mass
associated with the Einstein ring of the lenses has the form (Gould
1995; Turner, Ostriker \& Gott 1984):
\beq \sigma_{\rm
lens}(z_{\rm L})=\frac{4\pi G}{c^2} \frac{x_{\rm L}(\lambda_{\rm S}-
\lambda_{\rm L})\lambda_{\rm L}}{\lambda_{\rm S}}.
\label{eq:sigmalens}
\eeq
Here the subscripts L and S refer to the lens and the source,
$x\equiv (1+z)$, and
\beq
\lambda\equiv\frac{2}{5}\frac{c}{H_0}
\left({1-x^{-5/2}}\right).
\label{eq:gottdef}
\eeq
We implicitly assume that the intergalactic star clusters are
intrinsically optically thin to microlensing.

\noindent
{\it Optical Depth to Electron Scattering}

The optical depth to electron scattering out to a redshift $z$ in an
$\Omega_0=1$ universe, is given by,
\beq
\tau_{\rm es}(z)=0.053\Omega_{\rm b} h\int_0^{z} dz^\prime
\sqrt{1+z^\prime} \left[1-f_{\rm star} F_{\rm coll}(z^\prime)\right]
F_{\rm HII}(z^\prime),
\label{eq:taues}
\eeq
where we used the fact that the average ionization fraction is equal to the
HII filling factor.  We have implicitly assumed that the fraction of
singly ionized helium is $F_{\rm HII}$, and that no helium is ionized
twice due to the sharp drop in our template spectrum  
at the HeII edge (cf. Fig.~\ref{fig:spectrum}).

\noindent
{\it CMB Anisotropy Damping}

The CMB anisotropies are damped on scales smaller than the apparent
angular size of the horizon at the reionization epoch ($\sim
10^\circ$), due to scattering off free electrons.  Hu \& White (1996)
provide a fitting formula for the damping factor of the CMB
power--spectrum, $R_l^2$, as a function of the index $l$ in the
spherical harmonic decomposition of the microwave sky [the angular
scale corresponding to a given $l$--mode is
$\sim1^\circ\times(l/200)^{-1}$]. The damping factor is uniquely
related to the scattering optical depth as a function of redshift,
$\tau_{\rm es}(z)$, which is given by equation~(\ref{eq:taues})
(see also Kamionkowski et al. 1994).

\begin{table}[b]
\caption{\label{tab:models} 
The assumed parameter values in our standard model and its variants.}
\vspace{0.3cm}
\begin{center}
\begin{tabular}{|c||c|c|}
\hline
Parameter & Standard Model  & Range Considered \\
\hline
\hline  
$\sigma_{8h^{-1}}$   & 0.67             & 0.67--1.0 \\
\hline  
$n$                  & 1.0              & 0.8--1.0  \\
\hline
$\Omega_{\rm b}$     & 0.05             & 0.01--0.1 \\
\hline
$f_{\rm star}$       & 13\%              & 1\%--40\% \\
\hline
$f_{\rm esc}$        & $f_{\rm esc}(z)$ & 3\%--100\%  \\
\hline
IMF tilt ($\beta$)   & 0                & 0--1.69   \\
\hline
${\rm H_2}$ feedback & yes              & yes/no \\
\hline  
\end{tabular}
\end{center}
\end{table}

\section{Results and Discussion} 

\begin{figure}[b]
\vspace{2.6cm}
\includegraphics{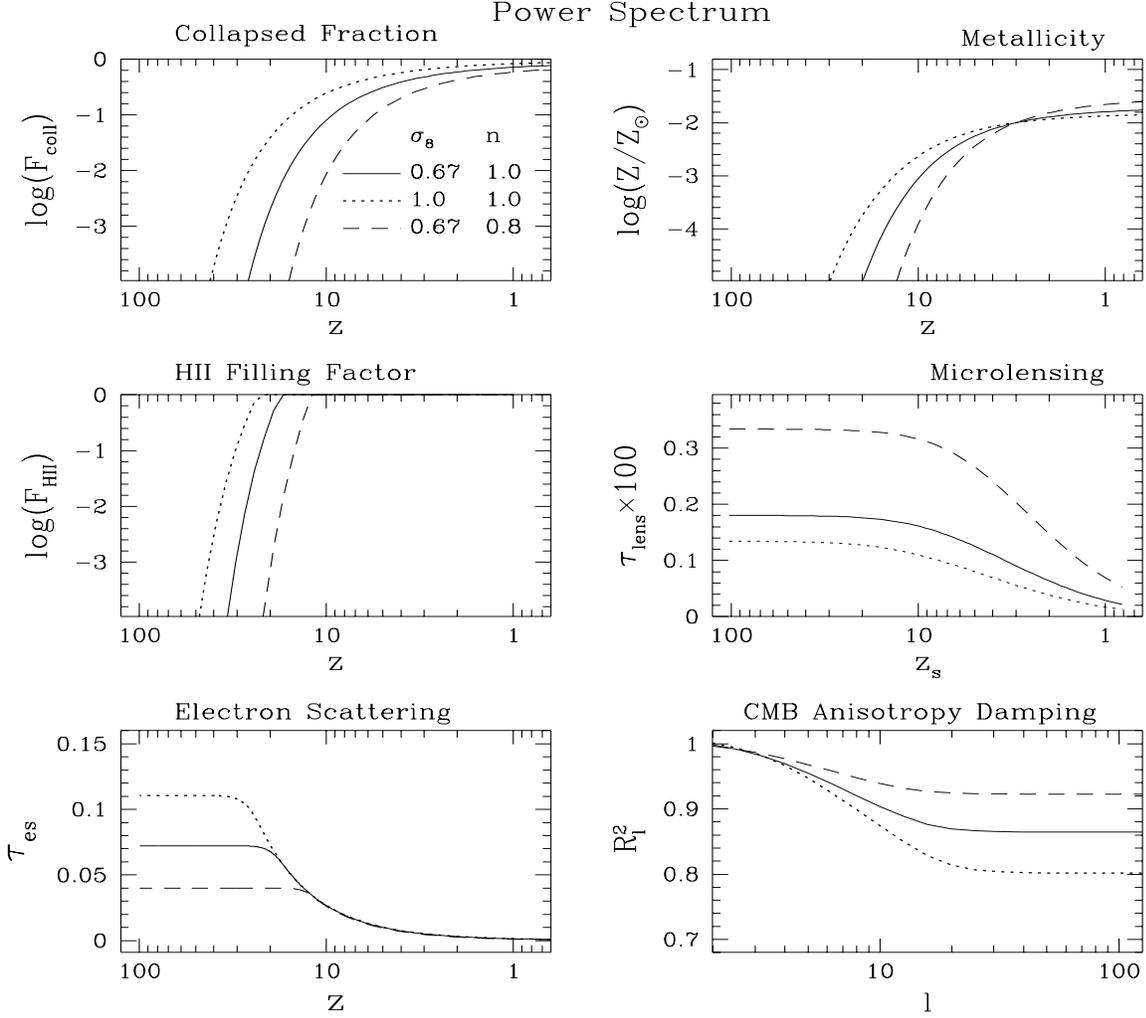}
\vspace*{4.5in}
\caption[IMF] {\label{fig:powspec} The different panels show (i) the collapsed
fraction of baryons; (ii) the average metallicity of the IGM; (iii) the
volume filling factor of ionized regions; (iv) the optical depth to
microlensing; (v) the optical depth to electron scattering; and (vi) the
net damping factor for the power--spectrum decomposition of CMB
anisotropies as a function of the spherical harmonic index $l$.  In this
figure we examine the sensitivity of the results to changes in the
normalization $\sigma_{8h^{-1}}$ or the primordial index $n$ of the CDM
power--spectrum.}
\end{figure}

\begin{figure}[b]
\vspace{2.6cm}
\includegraphics{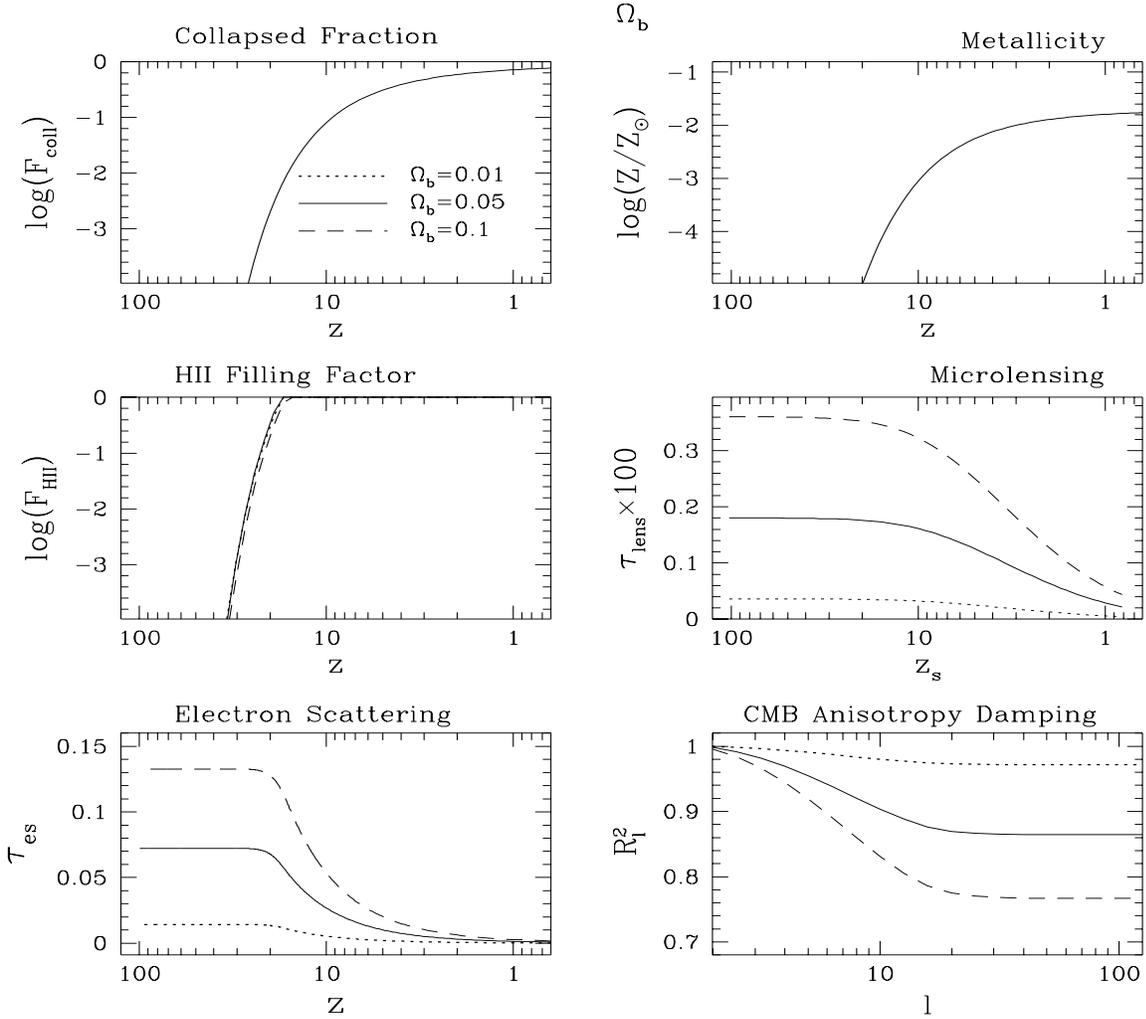}
\vspace*{4.5in}
\caption[IMF] {\label{fig:omegab} Same as Figure~\ref{fig:powspec}, but
showing the sensitivity of the results to different values of the
baryon density parameter,
$\Omega_{\rm b}$. }
\end{figure}

\begin{figure}[b]
\vspace{2.6cm}
\includegraphics{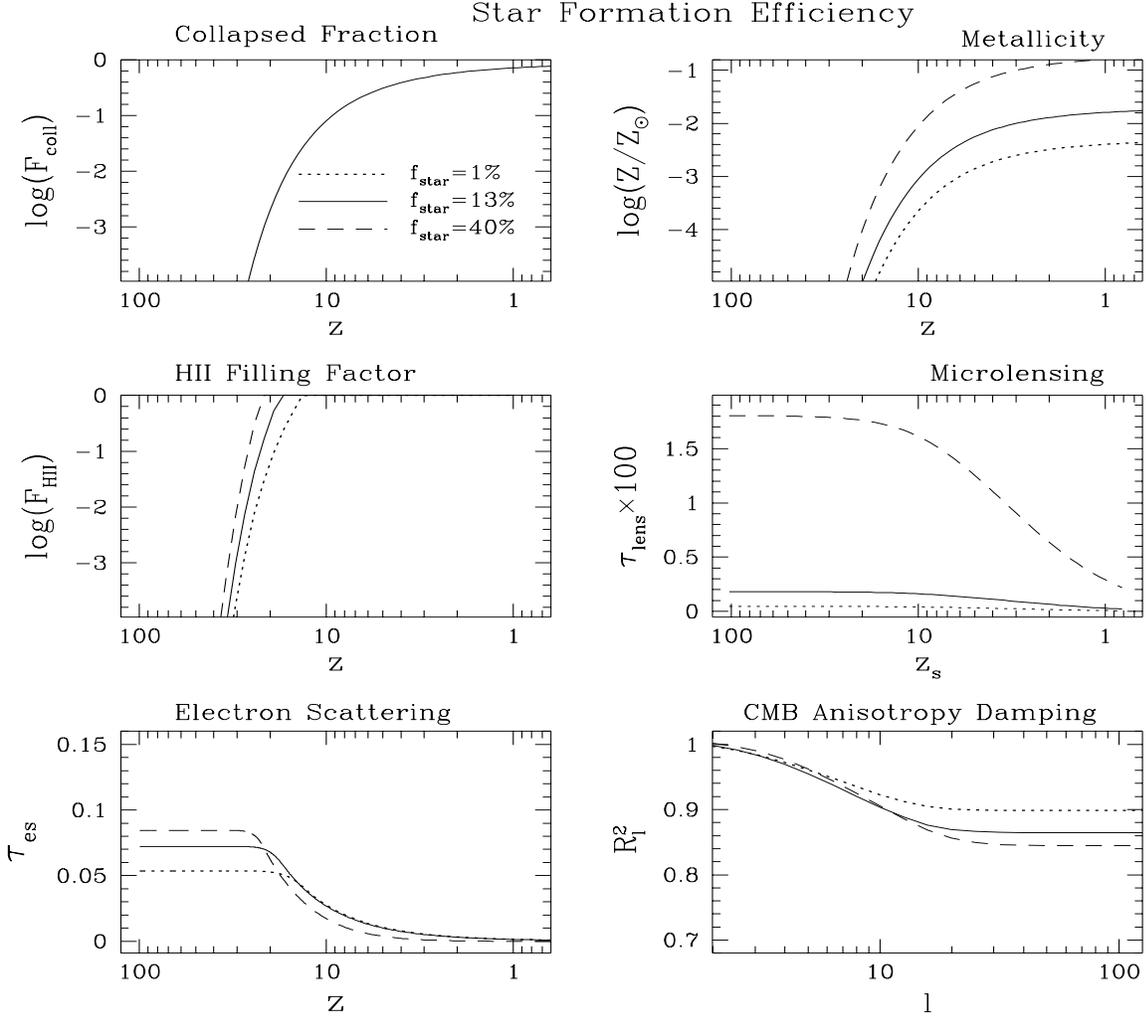}
\vspace*{4.5in}
\caption[IMF] {\label{fig:fstar} Same as Figure~\ref{fig:powspec}, but
showing the sensitivity of the results to a change in the star formation
efficiency $f_{\rm star}$.  If the carbon output from the stars is kept
fixed, then a change in $f_{\rm star}$ translates to a proportional change
in the metallicity, as shown in the upper right panel.  However, the carbon
yield from stars is uncertain, and thus for the other panels--an increase of
$f_{\rm star}$ by some factor may also be viewed as a decrease
in the carbon yield by the same factor,
while keeping the IGM metallicity fixed.}
\end{figure}

\begin{figure}[b]
\vspace{2.6cm}
\includegraphics{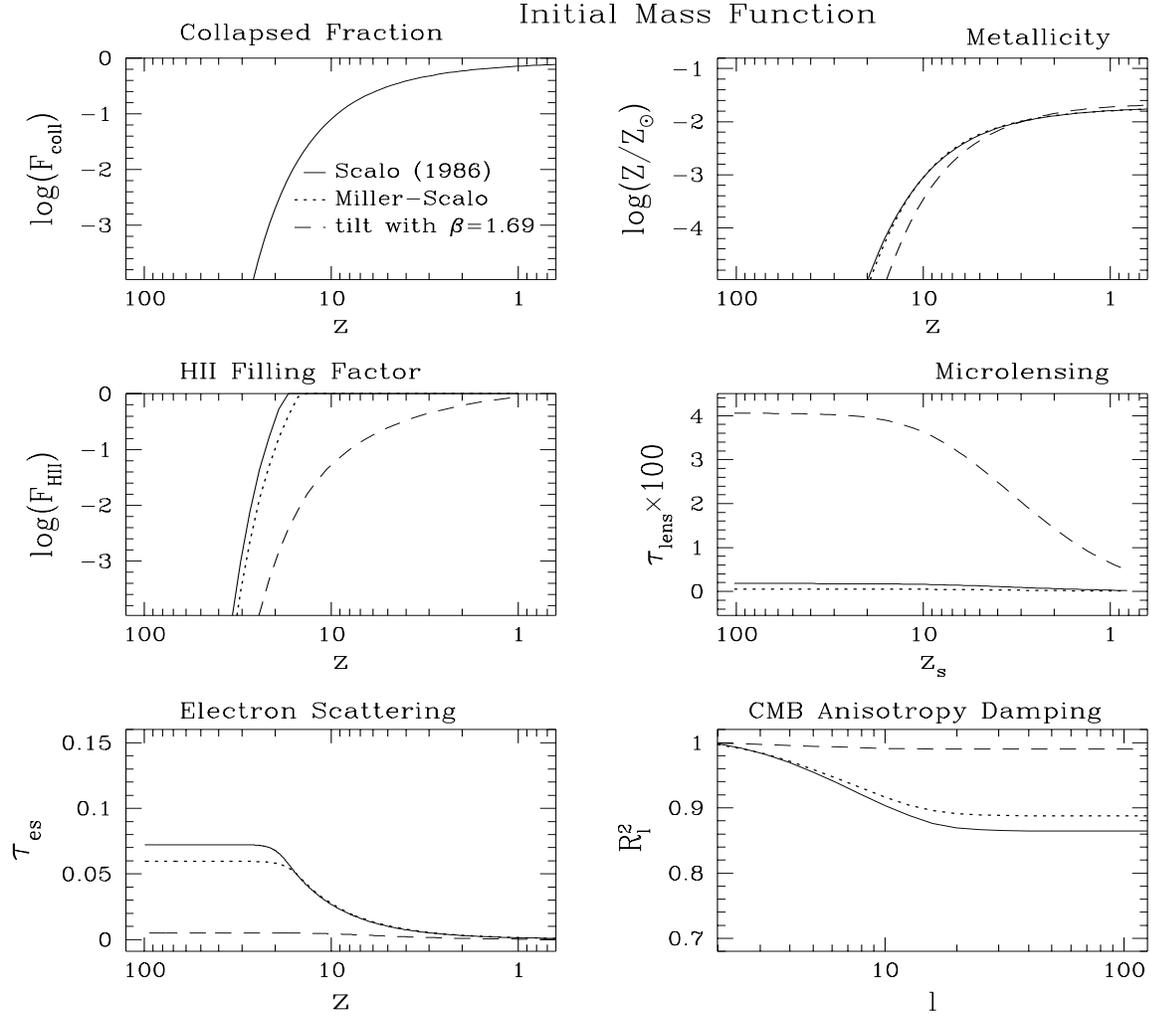}
\vspace*{4.5in}
\caption[IMF] {\label{fig:tilt} Same as Figure~\ref{fig:powspec}, but
showing the effect of tilting the Scalo IMF, i.e. making it steeper by
adding a constant $\beta$=1.69 to the power law index in each segment.
Results are also shown for a Miller--Scalo IMF which shows the
opposite bias in that it contains more high mass stars than the Scalo
IMF.}
\end{figure}

\begin{figure}[b]
\vspace{2.6cm}
\includegraphics{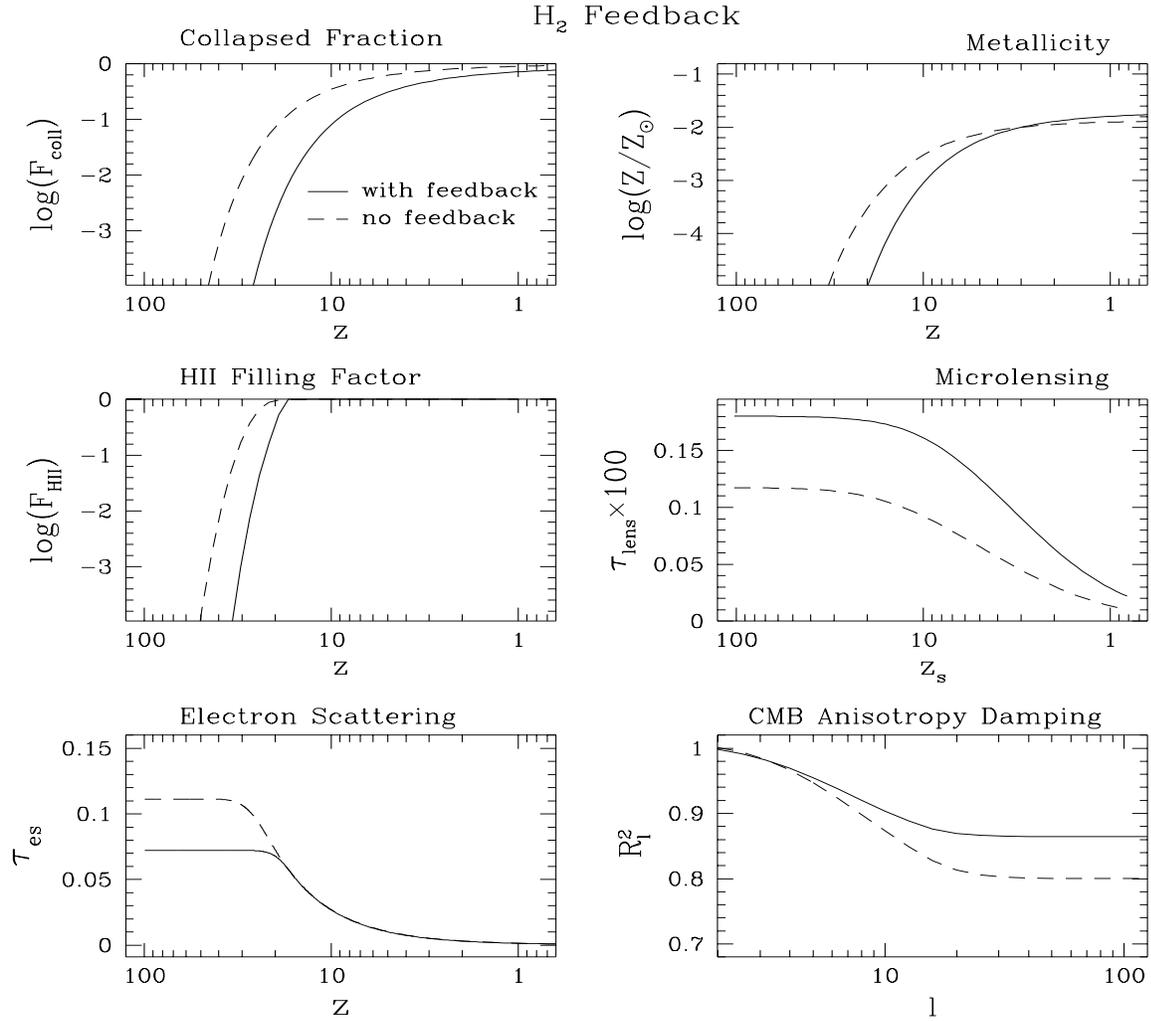}
\vspace*{4.5in}
\caption[IMF] {\label{fig:h2feedback} Same as Figure~\ref{fig:powspec}, but
showing the significance of the negative feedback on star formation
due to the photodissociation of ${\rm H_2}$.}
\end{figure}

\begin{figure}[b]
\vspace{2.6cm}
\includegraphics{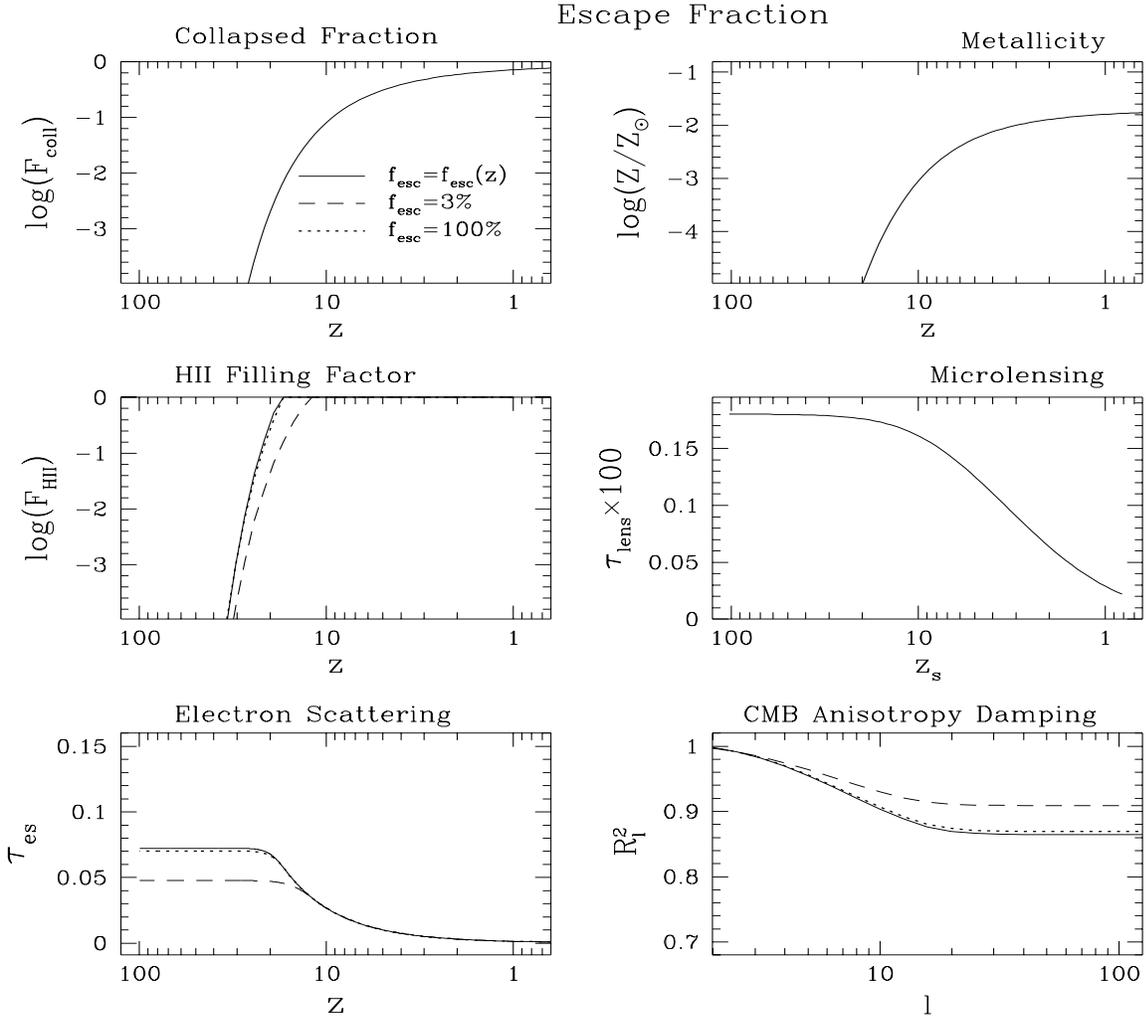}
\vspace*{4.5in}
\caption[IMF] {\label{fig:fesc} Same as Figure~\ref{fig:powspec}, but
showing the sensitivity of the results to changes in the the escape
fraction of ionizing photons, $f_{\rm esc}$.}
\end{figure}

Figures~\ref{fig:powspec}--\ref{fig:fesc} show results from our
reionization models.  We define our ``standard model'' by the set of
parameters summarized in Table~\ref{tab:models}, and analyze the
sensitivity of the results to separate changes in the values of each
parameter. Our choice for the cosmological parameters follows standard
CDM with a scale--invariant power spectrum, and the likely
nucleosynthesis value for the baryon density (Copi, Schramm \& Turner
1995), namely $n$=1, $h$=0.5, $\Omega=1$, $\Omega_{\rm b}$=0.05.  The
COBE normalization of this power spectrum implies
$\sigma_{8h^{-1}}$=1.22, which predicts too much power on small scales
and is ruled out by observations (Bunn \& White 1996). We therefore
use as our standard value $\sigma_{8h^{-1}}$=0.67 (an upper limit
based on cluster abundance, see White, Efstathiou, \& Frenk 1993; Bond
\& Myers 1996; Viana \& Liddle 1996). It is important to note that our
results depend primarily on the behavior of the power spectrum on
small scales and have little sensitivity to its characteristics on
large scales, where direct constraints from galaxy surveys and
microwave anisotropy experiments apply.  In our standard model, we
adopt a Scalo stellar IMF with the maximum carbon output (i.e. the
minimum amount of ionizing photons for a given metallicity), and
assume the existence of a negative feedback on star--formation due to
photodissociation of ${\rm H_2}$. The standard star formation
efficiency $f_{\rm star}$ is chosen to be 13\%, so as to yield a
metallicity of $0.01Z_{\odot}$ at a redshift $z=3$, where the
collapsed fraction is $F_{\rm coll}\approx 50\%$ (cf. \S 2.4).  The
standard escape fraction $f_{\rm esc}$ is explicitly calculated at
each redshift as described in \S~2.2 and shown in
Figure~\ref{fig:escape}.

The results from our standard model are described by the solid lines
in Figures~\ref{fig:powspec}--\ref{fig:fesc}.  In this model,
reionization occurs at $z=18$, with a corresponding scattering optical
depth of $\tau_{\rm es}=0.07$.  This leads to a $\sim 14\%$ reduction
in the power of CMB anisotropies on angular scales $\la 10^\circ$. The
microlensing optical depth is small, $\tau_{\rm lens}\la
2\times10^{-3}$, even at high redshifts.  Figure~\ref{fig:powspec}
also shows results for a higher normalization ($\sigma_{8h^{-1}}$=1.0)
and a tilt ($n$=0.8) of the CDM power spectrum.  With these
variations, the reionization redshift changes to $z$=22 and $z$=13,
and the scattering optical depth changes to $\tau_{\rm es}=0.11$ and
$\tau_{\rm es}=0.04$, respectively.  The low level of these changes
demonstrates that our results are robust for popular variants of the
CDM power spectrum.

Next, we consider the dependence of the results on the baryon density.
Figure~\ref{fig:omegab} shows the cases of $\Omega_{\rm b}$=0.01 and
0.1.  The microlensing optical depth is simply proportional to
$\Omega_{\rm b}$, while the electron scattering optical depth for
these cases is $\tau_{\rm es}=0.02$ and $\tau_{\rm es}=0.13$,
respectively.  The deviation from the exact proportionality $\tau_{\rm
es} \propto \Omega_{\rm b}$ is due to the increased recombination rate
in a high baryon density universe, which slightly decreases the
reionization redshift.

Figure~\ref{fig:fstar} illustrates the significance of changing the
star formation efficiency $f_{\rm star}$.  Although $f_{\rm star}$ is
directly constrained by the observed IGM metallicity, there are two
reasons to consider variations in this parameter.  First, there is
approximately an order of magnitude uncertainty in the observed value
of the CIV/H ratio; this makes the metallicity and therefore $f_{\rm
star}$ uncertain by the same factor.  Second, if hot bottom burning is
efficient, it could eliminate the carbon yields from 3--8${M_{\odot}}$
stars, and increase $f_{\rm star}$ by up to a factor of 4.  We
therefore examine the cases $f_{\rm star}$=1\% and $f_{\rm
star}$=40\%.

For $f_{\rm star}$=40\% the reionization redshift is increased to
$z$=24, but (i) the reionization in this case is more sudden; and (ii)
a substantial fraction of the electrons are locked up in stars and do
not contribute to $\tau_{\rm es}$ (the gas fraction is reduced by the
factor $[1-f_{\rm star}F_{\rm coll}]$, cf.~eq.~[\ref{eq:taues}]).
These effects counteract the increase in the reionization redshift,
and yield an optical depth, $\tau_{\rm es}=0.09$, which is only 15\%
higher than with $f_{\rm star}$=13\%.  On the other hand, $f_{\rm
star}$=1\% gives $\tau_{\rm es}$=0.05, again not too different from
$\tau_{\rm es}$=0.07.  Hence, Figure~\ref{fig:fstar} demonstrates that
the scattering optical depth is fairly insensitive to large variations
in the carbon output.

Figure~\ref{fig:tilt} demonstrates the effect of varying the IMF.
Although the stellar IMF at high redshift is unknown, it is often
claimed that due to the absence of metals and the lack of initial
substructure, it favored massive stars (e.g. Carr, Bond, \& Arnett
1984).  For this reason, we show in Figure~\ref{fig:tilt} the results
for the Miller--Scalo IMF (dotted lines), which is biased towards
massive stars relative to the standard Scalo IMF (solid lines).  Note
that the bias towards massive stars increases both the carbon yield
and the UV photon production per solar mass.  Therefore, a smaller
number of stars are necessary to provide the observed metallicity
($f_{\rm star}=5$\%), but these stars produce more ionizing photons.
As the figure shows, the two effects almost cancel each other, and
reionization with the Miller--Scalo IMF occurs at $z=14$, close to the
redshift with the Scalo IMF.

On the other hand, the possibility still exists that the
high--redshift IMF is biased in the opposite sense, i.e. towards
low--mass stars.  Indeed, in order to identify the early stars with
MACHOs which contribute significantly to the mass of galaxy halos, a
substantial fraction of the baryons must be converted into stars, and
most of these stars must have low masses, $\lsim0.5{\rm M_{\odot}}$
(Flynn et al. 1996; Alcock et al.  1996).  We therefore assume $f_{\rm
star}=90\%$. The tilt necessary to increase $f_{\rm star}$ from 13\%
to 90\% while keeping the metallicity fixed, must decrease the total
carbon yield per unit mass by a factor of 60. This is achieved by
adding a constant $\beta=1.69$ to the power--law index in all segments
of the Scalo IMF. This tilt decreases the average stellar mass from
$0.5{\rm M_{\odot}}$ to $0.25{\rm M_{\odot}}$ (cf.
Fig.~\ref{fig:imf}).  The dashed lines in figure~\ref{fig:tilt} shows
that with this tilt, stellar reionization is delayed substantially
down to $z\sim 1$ (at which point quasars, which are ignored in this
work, dominate the ionizing background).  This delay is caused by the
lack of ionizing photons from massive stars and is therefore a robust
conclusion; combining a tilt with $f_{\rm esc}=100\%$ only increases
the reionization redshift up to $z\sim2$.

According to Figure~\ref{fig:tilt}, a tilted IMF increases the optical
depth to microlensing substantially up to values as high as 3\% at
$z$=5 or 4\% at $z\gsim30$. Such microlensing probabilities could
leave detectable signatures in large samples of high redshift quasars.
Microlensing could either be detected through measurements of the
variability of quasars or the systematic changes in the equivalent
width of their emission lines.  The variability signature requires an
impractically long observing period (of order tens of years for solar
mass lenses), and is difficult to separate from an intrinsic
variability signal.  The equivalent--width change results from the
fact that a stellar lens could magnify the quasar continuum source,
which is typically smaller than its Einstein radius $\sim 10^{16}
(M_\star/0.1M_\odot)^{1/2}~{\rm cm}$, but would not magnify the broad
line region which is much more extended (see Dalcanton et al.  1994,
and references therein).

Figure~\ref{fig:h2feedback} shows the results with and without the
${\rm H_2}$-photodissociation feedback on star-formation.  Even though
the collapsed baryonic fraction is much higher without this feedback,
in both cases reionization occurs at about the same redshift,
$z\approx 20$.  The similarity in the reionization redshifts results
from the fact that when the HII filling factor approaches unity, star
formation is suppressed both with and without the ${\rm H_2}$ feedback
(cf. \S~3). The total optical depth to electron scattering without the
${\rm H_2}$--feedback is increased to $\tau_{\rm es}=0.11$; most of
the difference from the case with the ${\rm H_2}$--feedback is
introduced when the universe is only partially ionized.

Figure~\ref{fig:fesc} illustrates the sensitivity of the results to
the escape fraction.  We show results for either the low value of
$f_{\rm esc}$=3\% observed in nearby starburst galaxies (Leitherer
et~al.~1996), or the hypothetical high value of 100\%.  The motivation
for considering $f_{\rm esc}$=100\% is the theoretical possibility
that the obscuring gas could be expelled from the star forming clouds
by photoionization heating or supernova--driven winds.  Such an
expulsion of the gas would then quench further star-formation in these
systems (Lin \& Murray 1992); this mechanism could explain why the
star formation efficiency is limited to 13\%.  In this work, however,
we normalize the star formation efficiency based on the metallicity of
the \lya~forest, without attempting to explain its origin.  In the
case of a low escape fraction, the resulting optical depth for
electron scattering is decreased to $\tau_{\rm es}$=0.05, with
reionization still occurring at $z$=11. In the case of a high escape
fraction, the optical depth and reionization redshifts are increased
by a negligible fraction.

\begin{figure}[b]
\vspace{2.6cm}
\includegraphics{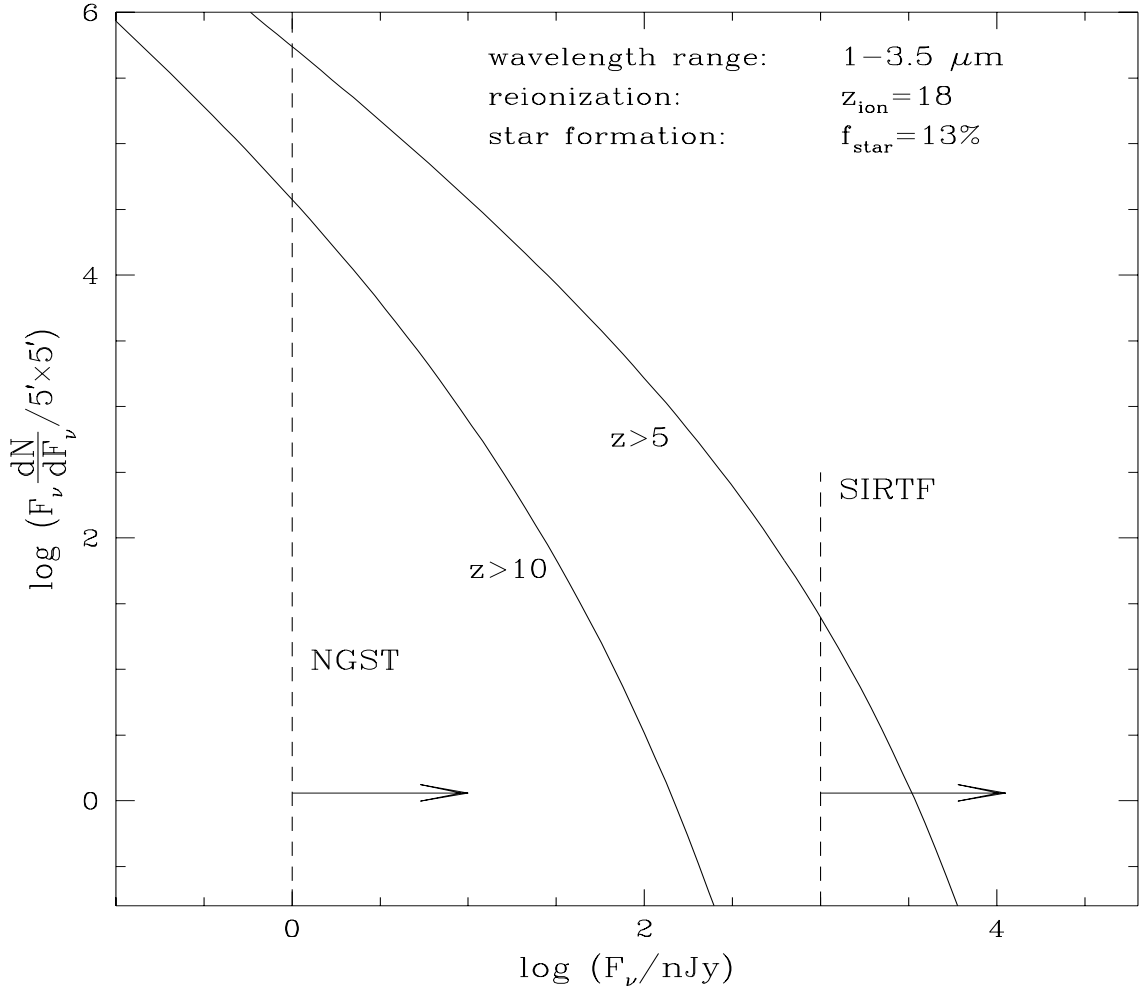}
\vspace*{4.5in}
\caption[NGST LF] {\label{fig:ngst} The number per 
logarithmic flux interval of high--redshift objects which could be probed
by future space telescopes, in the wavelength range of 1--3.5$\mu$m.  In
this calculation, we assume sudden reionization at $z$=25.  The vertical
dashed lines show the expected sensitivities of the Space Infrared
Telescope Facility (SIRTF) and the Next Generation Space
Telescope (NGST).}
\end{figure}

In addition to the above indirect signatures of reionization, it is
interesting to examine the possibility of a direct imaging of the
reionization sources.  Although any such imaging is beyond the
sensitivities of current instruments, the Next Generation Space
Telescope (NGST) will achieve a suitable sensitivity for that purpose,
$\sim 1$ nJy in the wavelength range 1--3.5$\mu$m (Mather \& Stockman
1996; see also http://ngst.gsfc.nasa.gov).  In Figure~\ref{fig:ngst}
we show the predicted number of objects that NGST would detect as a
function of observed flux, assuming a sudden reionization at the
redshift $z_{\rm ion}=18$.  The number counts are normalized to the
field of view of the Space Infrared Telescope Facility (SIRTF)
$5^{\prime}\times5^{\prime}$, which is scheduled for launch long
before NGST (the planned field-of-view of NGST is similar,
$4^\prime\times4^\prime$).  The abundance and emissivity of each
source was calculated according to our standard model as described in
\S~2.1-2.2.  We have averaged the flux from each source over the range
of 1--3.5$\mu$m and truncated the spectra of all objects at emission
frequencies above \lya~before reionization occurs, since the optical
depth of HI at these frequencies is exceedingly high prior to
reionization.

Figure~\ref{fig:ngst} shows the number per logarithmic flux interval
of all objects at $z>5$ (top curve), and all objects at $z>10$ (bottom
curve).  The number of detectable sources is high. With its expected
sensitivity, NGST will be able to probe about $10^4$ objects at $z>5$.
The average separation between these compact sources would be of order
2.4\H{}, well above the angular resolution of 0.06\H{}, planned for
NGST.

Another detectable signature of a reionized IGM results from its
Bremsstrahlung emission.  In an $\Omega=1$ universe, the surface
brightness observed today is given by (Loeb 1996)
\beq
J_{\rm ff}(\nu)=\frac{c}{H_0}\int_{\infty}^{0}\frac{j_{\rm ff}[(1+z)\nu]dz}
{(1+z)^{11/2}},
\eeq
where the free--free emissivity $j_{\rm ff}$ at very low frequencies is
independent of frequency (we ignore the logarithmic
frequency--dependence of the Gaunt factor), and is given by (Rybicki
\& Lightman 1979)
\beq
j_{\rm ff}(\nu)= 3.1\times 10^{-40} \frac{\langle n_{\rm e}^2\rangle}
{(T/10^4{\rm K})^{0.5}}~~~{\rm {erg~cm^{-3}~s^{-1}~Hz^{-1}~sr^{-1}}}.
\eeq
The value of $\langle n_{\rm e}^2\rangle$ at each redshift is a sum of the
contributions from the ionized regions of the smooth IGM 
and the star--forming clouds themselves,
\beq
\langle n_{\rm e}^2\rangle=F_{\rm HII}(1-F_{\rm coll})
\langle n_{\rm e}^2\rangle_{\rm IGM}
+\int_{\infty}^{z}dz^{\prime}\frac{dF_{\rm coll}}{dz}(z^{\prime})
\langle n_{\rm e}^2\rangle_{_\star}[t(z)-t(z^{\prime})].
\eeq
In our standard model, the average value of $n_{\rm e}^2$ within an ionized
region of the smooth IGM is simply given by $\langle n_{\rm
e}^2\rangle_{\rm IGM}=[1.15\times10^{-7}(1+z)^3]^2~~{\rm cm}^{-6}$. The
average of $n_{\rm e}^2$ within each star forming cloud can be obtained by
noticing that the recombination rate in each cloud is simply $(1-f_{\rm
esc})$ times the production rate of ionizing photons, i.e.
\beq
\langle n_{\rm e}^2\rangle_{_\star}[\Delta t] \equiv 
\frac{1}{\frac{4}{3}\pi R^3} \int_0^R 4\pi r^2 dr n_{\rm e}^2(r) = 
\frac{1}{\alpha_{\rm B}} \frac{1}{\frac{4}{3}\pi R^3} 
\left({1-f_{\rm esc}\over f_{\rm esc}}\right)
\frac{dN_{\gamma}}{dt}[\Delta t] .
\eeq
This result for the clumpiness factor is independent of the
cloud mass and is only a function of the time lag since 
the starburst occurred, $\Delta t$. 
The escape fraction $f_{\rm esc}$ in this expression was derived in \S~2.2
(cf.~Fig.~\ref{fig:escape}), and the time--dependent production rate of 
ionizing photons $dN_{\gamma}/dt$ was calculated in \S~2.3
(cf.~eq.~\ref{eq:source} and Fig.~\ref{fig:flux}).  

\begin{figure}[b]
\vspace{2.6cm}
\includegraphics{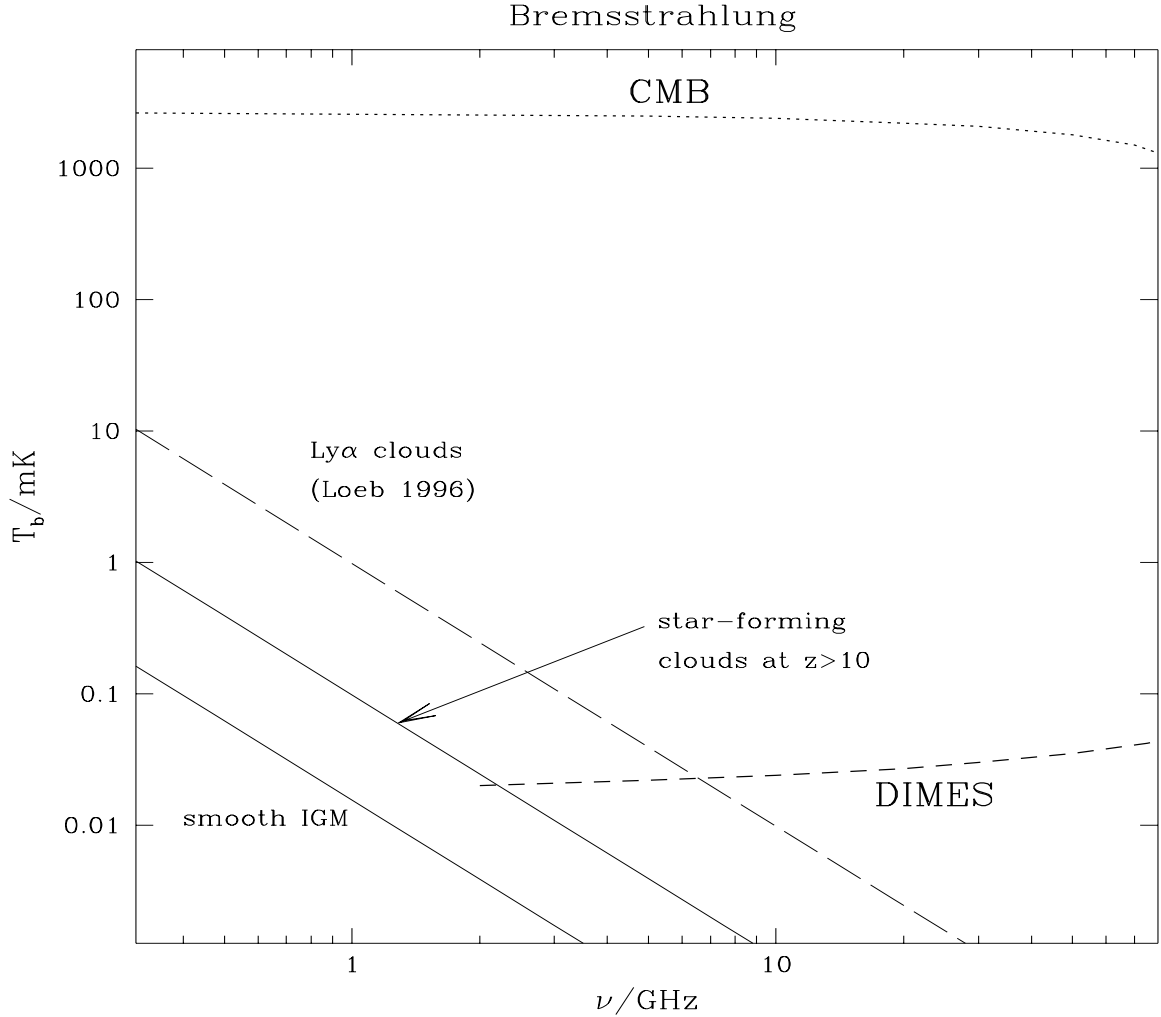}
\vspace*{4.5in}
\caption[Dimes] {\label{fig:brem} Observed brightness temperature
from Bremsstrahlung emission by
star--forming clouds at $z>10$ and ionized regions of the smooth IGM 
in our standard model.  The emission from the
star--forming clouds exists only if the clouds retain their gas over the
lifetime of their stellar population.  For comparison, we show the
Bremsstrahlung emission from the \lya~forest between $0<z<5$ for a UV
background of $10^{-21}~{\rm erg~cm^{-2}s^{-1} Hz^{-1}sr^{-1}}$ at the
Lyman--limit (Loeb~1996).  We also show the CMB spectrum, and the proposed
sensitivity of the DIMES experiment (Kogut~1996; see also
http://ceylon.gsfc.nasa.gov/DIMES).}
\end{figure}

The resulting brightness temperature, $T_{\rm b}\equiv (c^2 J_{\rm
ff}/ 2k \nu^2)$, of the microwave sky due to Bremsstrahlung emission
by star--forming clouds at $z>10$ and the smooth IGM, is shown in
Figure~\ref{fig:brem} for the frequency range $\nu=1$--$100$ GHz.  The
existence of Bremsstrahlung emission from star forming regions relies
on their ability to retain their gas during the lifetime of their
stellar population, and is therefore uncertain. Reprocessing of
starlight by dust (Wright 1981, Wright et al. 1994) might add to this
emission component and will be considered elsewhere (Loeb \& Haiman
1997).  For comparison, the long--dashed curve shows the unavoidable
Bremsstrahlung signal from the
\lya~forest clouds (with HI column densities $\la 10^{17}~{\rm cm^{-2}}$)
at $0\la z\la 5$, for a UV background flux of $10^{-21}~{\rm
erg~cm^{-2}s^{-1} Hz^{-1}sr^{-1}}$ at the Lyman--limit (Loeb~1996).  The
calculation of the \lya~forest contribution avoids any uncertainties about
the clumpiness of the IGM, because at photoionization equilibrium the
line-of-sight integral of $n_e^2$ is proportional to the {\it observed}
column density of neutral hydrogen.  The above signals are compared to the
expected sensitivity of the proposed Diffuse Microwave Emission Survey
(DIMES) experiment (dashed curve, from Kogut~1996; see also
http://ceylon.gsfc.nasa.gov/DIMES).  The (uncertain) contribution from the
star--forming clouds is comparable to the galactic free-free emission and
can easily be detected by the DIMES experiment.  Extreme variants of our
standard model (e.g., with a larger $\Omega_b$ or a lower $f_{\rm esc}$)
which enhance $J_{\rm ff}$ by more than an order of magnitude, might
already be ruled--out by the existing COBE limits on the spectral
distortion of the CMB (Mather et al.  1994; Wright et al. 1994). The
unavoidable contribution of \lya~clouds at $z\la 5$ (Loeb~1996) is
detectable with the future DIMES experiment, but the emission from the
smooth IGM at $z\ga 10$ is just below the sensitivity of this instrument in
our standard model.

\section{Conclusions}

For a large range of CDM models with a Scalo IMF, the universe is
reionized by a redshift $\ga 10$.  The optical depth to electron
scattering due to reionization is in the range $\sim 0.05$--$0.15$.
The resulting damping amplitude of $\sim 10$--$20\%$ for microwave
anisotropies on angular scales $\la10^\circ$, will be detectable with
future satellite experiments such as MAP or PLANCK.  These experiments
might be able to isolate a damping amplitude of several percent if
they measure the microwave polarization (Zaldarriaga 1996).

If the stellar IMF is strongly tilted relative to the local Scalo IMF
so that most of the baryons are transformed into low mass stars, then
stellar reionization is suppressed due to the absence of massive stars
which ordinarily dominate the ionizing flux (cf. Fig.~\ref{fig:flux}).
In this case, however, the existence of primordial stars might be
inferred from the measurable microlensing event rate that they produce
in the halo of the Milky Way (Alcock et~al.~1996; Flynn, Gould \&
Bahcall 1996) or along the lines of sight to distant quasars (Gould
1995; Dalcanton et al. 1994).

Aside from the above mentioned sensitivity to a drastic variation in the
IMF, much of the qualitative results summarized in
Figures~\ref{fig:powspec}--\ref{fig:fesc} appear robust to changes in other
model parameters.  This robustness results from a number of reasons. The
collapsed fraction depends exponentially on redshift in the rare tail of
the Gaussian random field of density fluctuations--from where the first
objects are drawn.  The inverse dependence of the collapse redshift, and
hence the scattering optical depth $\tau_{\rm es}$, on the power spectrum
is therefore only logarithmic.  The linear scaling of $\tau_{\rm es}$ with
$\Omega_{\rm b}$ is somewhat moderated by the dependence of the
recombination rate on $\Omega_{\rm b}^2$.  The reionization redshift is not
sensitive to variations in the star formation efficiency $f_{\rm star}$ or
the escape fraction $f_{\rm esc}$ , because of the strong dependence of the
recombination rate on redshift ($\propto[1+z]^6$).  Finally, the
suppression of star--formation due to the H$_2$ feedback is significant
compared to the suppression due to photo--ionization heating only as long
as the filling factor of ionized bubbles is still small; once the ionized
bubbles start to overlap, the H$_2$ feedback becomes unimportant.

We have calibrated the metal production in our models based on the
carbon abundance detected in the \lya~forest at $z\sim3$
(Cowie~et~al.~1995; Tytler~et~al.~1995; Songaila \& Cowie 1996, Cowie
1996).  If the metallicity of the \lya~forest, as inferred from the
carbon/hydrogen ratio, is indeed universal and due to a single
generation of stars, then the star formation efficiency for a Scalo
IMF is higher than $4\%$, or else ``hot bottom burning'' (which
destroys carbon in stellar envelopes) is entirely inefficient.  This
conclusion is independent of the details of the rest of our
modeling. Future detection of other metal absorption lines in the
\lya~forest would allow to discriminate among these possibilities.

Much of the uncertainty in our reionization models results from the lack of
a standard theory for star formation.  Existing theoretical work on star
formation has been mainly guided by observations and therefore focused, by
and large, on the complex environments of the interstellar medium of our
galaxy.  The conditions in the early universe are much better defined than
those in the local interstellar medium, since the composition of the gas
and hence its chemistry and cooling function are simple and well known, the
power spectrum of initial density fluctuations is prescribed in specific
cosmological models, and magnetic fields are probably absent before stars
form. From a theoretical standpoint, the early universe offers a simpler
environment for understanding star formation. The interplay between theory
and observations of the reionization epoch might therefore add insight to
the physics of star formation, especially if the star clusters at high
redshifts are directly imaged. Our estimates, illustrated in
Figure~\ref{fig:ngst}, imply that deep infrared observations with the Next
Generation Space Telescope might be able to probe $\ga 10^3$ bright star
clusters at $z\sim5$--$10$.  The cumulative Bremsstrahlung emission from
these clusters (cf. Figure~\ref{fig:brem}) could potentially be detected by
future instruments such as the Diffuse Microwave Emission Survey (Kogut
1996).

\vspace{0.4cm} 

We thank George Field, Peter H\"oflich, Wayne Hu, Rohan Mahadevan, Martin
Rees, Dimitar Sasselov, David Spergel, and Martin White for stimulating
discussions and useful comments. AL acknowledges support from the NASA ATP
grant NAG5-3085.

\section*{REFERENCES}
{
\StartRef

\Ref Alcock, C., et al. 1996, preprint astro-ph/9606165

\Ref Arons, J., \& Wingert, D. W. 1972, ApJ, 177, 1

\Ref Bardeen, J.M., Bond, J.R., Kaiser, N., \& Szalay, A.S. 
1986, ApJ, 304, 15 (BBKS)

\Ref
Bechtold, J., Crotts, A. P. S., Duncan, R. C., 
\& Fang, Y. 1994, ApJ, 437, L83

\Ref Bertschinger, E. 1985, ApJS, 58, 39

\Ref Bond, J. R. 1995, in Theory and Observations of the Cosmic
Microwave Background Radiation, ed. Schaeffer, R. (Elsevier:
Netherlands), in press

\Ref Bond, J. R., \& Efstathiou, G. 1987, MNRAS, 226, 665

\Ref Bond, J. R., \& Myers, S. 1996, ApJS, 103, 63

\Ref Bunn, E. F., \& White, M. 1996, preprint astro-ph/9607060

\Ref Carr, B. J., Bond, J. R., \& Arnett, W. D. 1984, ApJ, 277, 445

\Ref Copi, C. J., Schramm, D. N., \& Turner, M. S. 1995, Science, 267, 192

\Ref Couchman, H. M. P. \& Rees, M. J. 1986, MNRAS, 221, 53

\Ref Cowie, L. 1996, in HST and the High Redshift Universe,
Proc. 37th Herstmonceux conference, eds. Tanvir, N. R., 
Arag\'on-Salamanca, A., Wall, J.V.  (World Scientific:Singapore), 
in press, (preprint astro-ph/9609158)

\Ref Cowie, L. L., Songaila, A., Kim, T.-S., \& Hu, E. M. 1995 AJ, 109, 1522

\Ref Dalcanton, J. J., Canizares, C. R., Granados, A., Steidel,
C. C., \& Stocke, J. T. 1994, ApJ, 424, 550

\Ref Davidsen, A. F., Kriss, G. A., \& Zheng, W. 1996, Nature, 380, 47

\Ref Dinshaw, N., Impey, C. D., Foltz, C. B., Weymann, R. J., 
\& Chaffee, F. H. 1994, ApJ, 437, L87

\Ref
Dinshaw, N., Foltz, C. B., Impey, C. D., Weymann, R. J., \& Morris, S. L.
1995, Nature, 373, 223
\Ref Dove, J. B., \& Shull, J. M. 1994, ApJ, 430, 222

\Ref Efstathiou, G., \& Bond, J. R. 1987, MNRAS, 227, 33p

\Ref Eisenstein, D.J., \& Loeb, A. 1995, ApJ, 443, 11

\Ref Fang, Y., Duncan, R. C, Crotts, A. P. S., \& Bechtold, J. 1995,
preprint astro-ph/9510112

\Ref Flynn, C., Gould, A., \& Bahcall, J. N. 1996, ApJ, 466, 55L

\Ref Fukugita, M., \& Kawasaki, M. 1994, MNRAS, 269, 563

\Ref Giroux, M. L., \& Shapiro, P. R. 1996, ApJS, 102, 191

\Ref Gnedin, N. Y., \& Ostriker, J. P. 1996, preprint astro-ph/9612127

\Ref Gould, A. 1995, ApJ, 455, 37

\Ref Haardt, F., Madau, P. 1996, preprint astro-ph/9609057

\Ref Haiman, Z., Thoul, A., \& Loeb, A. 1996, ApJ, 464, 523 (HTL96)

\Ref Haiman, Z., Rees, M. J., \& Loeb, A. 1996, ApJ, 467, 522 (HRL96)

\Ref Haiman, Z., Rees, M. J., \& Loeb, A. 1997, ApJ, in press (HRL97)

\Ref Hellsten, U., Dav\'e, R., Hernquist, L., Weinberg, D., \& Katz, N.
1997, preprint astro-ph/9701043

\Ref Hernquist, L., Katz, N., Weinberg, D. H., \& Miralda-Escud\'e, J.
1996, ApJ, 457, L51

\Ref Hu, W., \& White, M. 1996, ``The Damping Tail of CMB
Anisotropies'', preprint astro-ph/9609079

\Ref Kamionkowski, M., Spergel, D. N., \& Sugiyama, N. 1994, ApJL,
426, L57

\Ref Kogut, A. 1996, preprint astro-ph/9607100

\Ref Kurucz, R. 1993, CD-ROM No. 13, ATLAS9 Stellar Atmosphere Programs

\Ref Leitherer, C., Ferguson, H. C., Heckman, T. M., \& Lowenthal, J. D.
1995, ApJ, 454, L19

\Ref Lin D. C. \& Murray, S. D. 1992, ApJ, 394, 523

\Ref Loeb, A. 1996, ApJ, 459, L5

\Ref Loeb, A. \& Haiman, Z. 1997, in preparation

\Ref Madau, P., \& Shull, J. M. 1996, ApJ, 457, 551

\Ref Mather, J. C. et al. 1994, ApJ, 420, 439

\Ref Mather, J, \& Stockman, P. 1996, ST Sci Newsletter, v. 13, no. 2, p. 15

\Ref Meiksin, A., \& Madau, P. 1993, 412

\Ref Miller, G. E., \& Scalo, J. M. 1979, ApJS, 41, 513

\Ref Miralda-Escud\'e, J., \& Ostriker, J. P. 1990, ApJ, 350, 1

\Ref
Miralda-Escud\'e, J., Cen, R., Ostriker, J. P., \& Rauch, M. 1995,
ApJ, submitted, preprint astro-ph/9511013

\Ref Ostriker, J. P., \& Gnedin, Y. N. 1996, preprint astro-ph/9608047

\Ref Padmanabhan, T. 1993, Structure Formation in the Universe (Cambridge
University Press: Cambridge)

\Ref Press, W. H., \& Schechter, P. L. 1974, ApJ, 181, 425

\Ref Rauch, M., Haehnelt, M. G., \& Steinmetz, M. 1996, preprint 
astro-ph/9609083

\Ref Rees, M. J. 1996, preprint astro-ph/9608196

\Ref Renzini, A., \& Voli, M. 1981, A\&A, 94, 175

\Ref Rybicki, G. B., \& Lightman, A. P. 1979, Radiative Processes in 
Astrophysics (New York: Wiley)

\Ref Scalo, J. M. 1986, Fundamentals of Cosmic Physics, vol. 11, p. 1-278

\Ref Schaller, G., Schaerer, D., Meynet, G., \& Maeder, A. 1992, 
A\&ASS, 96, 269

\Ref Scott, D., Silk, J., \& White, M. 1995, Science, 268, 829

\Ref Shapiro, P. R., \& Giroux, M. L. 1987, ApJ, 321, L107

\Ref Shapiro, P. R., Giroux, M. L., \& Babul 1994, ApJ, 427, 25

\Ref Songaila, A., \& Cowie, L. L. 1996, AJ, 112, 335

\Ref
Smette, A., Robertson, J. G., Shaver, P. A., Reimers, D.,
Wisotzki, L., \& K\"ohler, Th. 1995, A\&A Suppl., 113, 199

\Ref Tegmark, M., Silk, J., \& Blanchard, A. 1995, ApJ, 434, 395

\Ref Tegmark, M., Silk, J., Rees, M. J., Blanchard, A., Abel, T.,
\& Palla, F. 1996, ApJ, in press, preprint astro-ph/9603007 

\Ref Turner, E. L. Ostriker, J. P., \& Gott, J., R., III. 1984, ApJ, 284, 1

\Ref Tytler, D. et al. 1995, in QSO Absorption Lines, ESO 
Astrophysics Symposia, ed. G. Meylan (Heidelberg: Springer), p.289

\Ref van den Hoek, L. B., \& Groenwegen, M. A. T. 1996, 
preprint astro-ph/9610030

\Ref Viana, P. T. P., \& Liddle, A. 1996, MNRAS, in press, 
preprint astro-ph/9511007

\Ref Vikhlinin, A. A. 1995, Astron. Lett., vol. 21, no. 3., p. 366

\Ref White, S. D. M., Efstathiou, G., \& Frenk, C. S. 1993, MNRAS, 262, 1023

\Ref Woosley, S. E., \& Weaver, T. A. 1986, ARA\&A, 24, 205 

\Ref Woosley, S. E., \& Weaver, T. A. 1995, A\&A Supp., 101, 181

\Ref Wright, E. L. 1981, ApJ, 250, 1 

\Ref Wright, E. L. et al. 1994, ApJ, 420, 450

\Ref Zaldarriaga, M. 1996, preprint astro-ph/9608050

}

\clearpage
\newpage

\clearpage
\newpage

\clearpage
\newpage

\clearpage
\newpage

\clearpage
\newpage

\clearpage
\newpage

\clearpage
\newpage

\clearpage
\newpage

\clearpage
\newpage
\clearpage
\newpage

\clearpage
\newpage

\clearpage
\newpage

\clearpage
\newpage

\clearpage
\newpage

\clearpage
\newpage

\clearpage
\newpage

\clearpage
\newpage

\end{document}